%% file: neurips.tex
\documentclass{article}

\PassOptionsToPackage{}{natbib}

\usepackage[preprint]{neurips_2021}

\usepackage[utf8]{inputenc} 
\usepackage[T1]{fontenc}    
\usepackage{hyperref}       
\usepackage{url}            
\usepackage{booktabs}       
\usepackage{amsfonts}       
\usepackage{nicefrac}       
\usepackage{microtype}      
\usepackage{multirow}

\usepackage[pdftex]{graphicx}
\usepackage{subfigure}
\usepackage{enumitem}
\setlist[itemize]{leftmargin=5.5mm}

\usepackage{url}            
\usepackage{braket} 
\usepackage{amsmath}
\usepackage{color}
\usepackage{mathtools}
\usepackage[toc,page]{appendix}
\usepackage{amsthm}
\usepackage{wrapfig}
\usepackage{float}
\usepackage{tcolorbox}
\usepackage{thmtools}
\usepackage{thm-restate}

\usepackage{siunitx}

\usepackage{pdfpages}

\newtheorem{proposition}{Proposition}

\newcommand{\x}{{\mathbf x}}

\newcommand{\z}{{\mathbf z}}

\newcommand{\h}{{\mathbf h}}

\newcommand{\bb}{{\mathbf b}}
\newcommand{\s}{{\mathbf s}}

\newcommand{\cc}{{\mathbf c}}

\newcommand{\e}{{\mathbf e}}

\newcommand{\A}{{\mathbf A}}

\newcommand{\I}{{\mathbf I}}
\newcommand{\HH}{{\mathbf H}}

\newcommand{\T}{{\mathbf T}}

\newcommand{\R}{{\mathbb R}}
\newcommand{\p}{{\mathbf p}}
\newcommand{\q}{{\mathbf q}}

\newcommand\myeq{\mathrel{\overset{\makebox[0pt]{\mbox{\normalfont\tiny\sffamily def}}}{=}}}

\usepackage{arydshln}

\usepackage{cleveref}

\makeatletter
\newcommand{\printfnsymbol}[1]{%
  \textsuperscript{\@fnsymbol{#1}}%
}
\newcommand{\specificthanks}[1]{\@fnsymbol{#1}}
\makeatother

\title{$\textsc{GeoMol}$: Torsional Geometric Generation of Molecular 3D Conformer Ensembles}

\author{
  Octavian-Eugen Ganea \thanks{Equal first authorship and contribution. Correspondence to \texttt{\{oct,lagnajit\}@mit.edu} .} \textsuperscript{ , \specificthanks{3}}
  \And
  Lagnajit Pattanaik\textsuperscript{ \specificthanks{1},} \thanks{Department of Chemical Engineering, MIT, Cambridge, MA 02139}
  \AND
  Connor W.~Coley \textsuperscript{\specificthanks{2}}
  \And
  Regina Barzilay \thanks{Computer Science and Artificial Intelligence Laboratory, MIT, Cambridge, MA 02139}
  \And
  Klavs F.~Jensen \textsuperscript{\specificthanks{2}}
  \AND
  William H.~Green \textsuperscript{\specificthanks{2}}
  \And
  Tommi S.~Jaakkola \textsuperscript{\specificthanks{3}}
}

\begin{document}

\maketitle

\begin{abstract}
 Prediction of a molecule's 3D conformer ensemble from the molecular graph holds a key role in areas of cheminformatics and drug discovery. Existing generative models have several drawbacks including lack of modeling important molecular geometry elements (e.g. torsion angles), separate optimization stages prone to error accumulation, and the need for structure fine-tuning based on approximate classical force-fields or computationally expensive methods such as metadynamics with approximate quantum mechanics calculations at each geometry. We propose \textsc{GeoMol}--an end-to-end, non-autoregressive and SE(3)-invariant machine learning approach to generate distributions of low-energy molecular 3D conformers. Leveraging the power of message passing neural networks (MPNNs) to capture local and global graph information, we predict local atomic 3D structures and torsion angles, avoiding unnecessary over-parameterization of the geometric degrees of freedom (e.g. one angle per non-terminal bond). Such local predictions suffice both for the training loss computation, as well as for the full deterministic conformer assembly (at test time). We devise a non-adversarial optimal transport based loss function to promote diverse conformer generation.  \textsc{GeoMol} predominantly outperforms popular open-source, commercial, or state-of-the-art machine learning (ML) models, while achieving significant speed-ups. We expect such differentiable 3D structure generators to significantly impact molecular modeling and related applications. \footnote{Code is available at \url{https://github.com/PattanaikL/GeoMol}.}
\end{abstract}


\input{intro}

\input{relwork}
\input{method}
\input{exp}
\section{Conclusion}
We proposed \textsc{GeoMol}, an end-to-end generative approach for molecular 3D conformation ensembles that explicitly models various molecular geometric aspects such as torsion angles or chirality. We expect that such differentiable structure generators will significantly impact small molecule conformer generation along with many related applications (e.g. protein-ligand binding), thus speeding up areas such as drug discovery. \textsc{GeoMol}'s full source code will be made publicly available.
 
\paragraph{Limitations \& future work. } A few current limitations are highlighted and left for future extensions (see also discussion in \cref{apx:add_discussion}). First, our model does not currently support  disconnected molecular graphs, e.g. ionic salts, but it can be applied to each connected component, followed by a 3D alignment. Next, our approach would benefit from explicit modeling of long distance interactions, especially for macrocycles or large molecules. This remains to be addressed in an efficient manner. Third, explicitly using  ground truth energy values could further improve \textsc{GeoMol}. Last, we look forward to fine-tune \textsc{GeoMol} on applications such as generating molecular docking poses or descriptors for 4D QSAR.


\begin{ack}
OEG thanks Bracha Laufer, Tian Xie, Xiang Fu, Peter Mikhael, and the rest of the RB and TJ group members for their helpful comments and suggestions. LP thanks Camille Bilodeau and the rest of the WHG, KFJ, and CWC research groups for their useful discussions. We also thank Pat Walters, Simon Axelrod, and Rafael Gomez-Bombarelli for their insightful feedback as well as Minkai Xu and Shitong Luo for helping with running the CGCF model. Both OEG and LP are funded by the Machine Learning for Pharmaceutical Discovery and Synthesis consortium.
\end{ack}

\bibliographystyle{plainnat}
\bibliography{biblio}

\newpage
\appendix
\input{appendix}

\end{document}

%% file: intro.tex
\begin{wrapfigure}{r}{0.4\textwidth}
\vspace{-1.8cm}
  \begin{center}
    \includegraphics[width=0.4\textwidth]{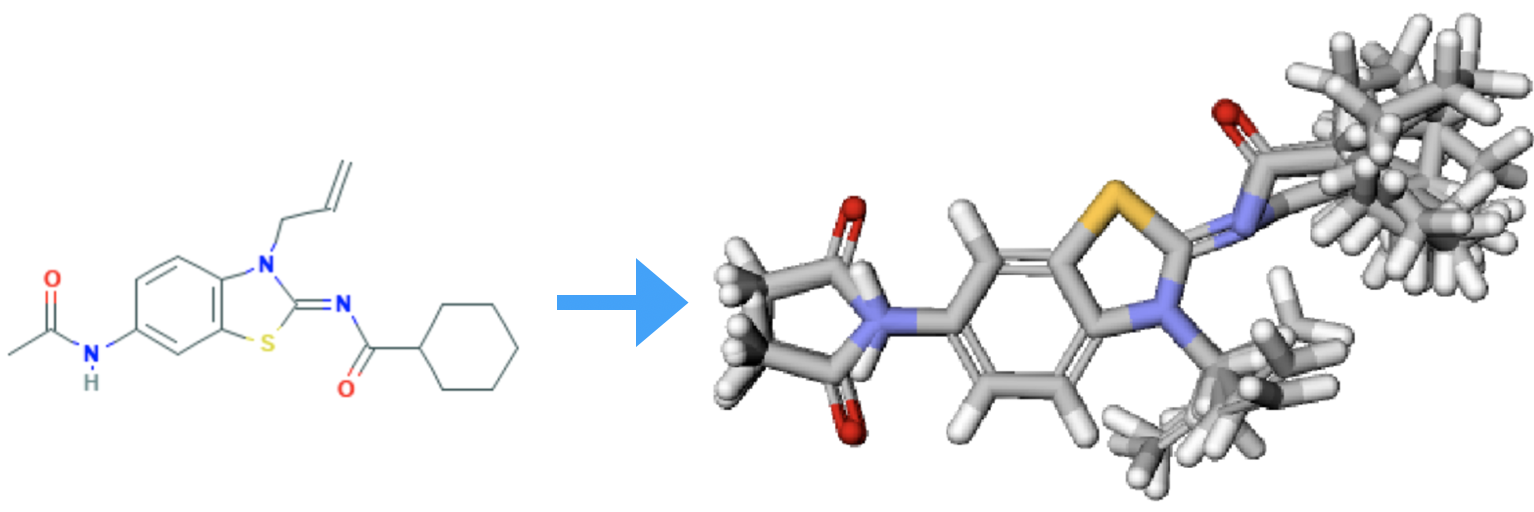}
  \end{center}
  \caption{We explore the task of low-energy 3D conformer ensemble generation from the input molecular graph. This example molecule has both rigid (rings) and more flexible parts. Conformers are shown aligned and juxtaposed. }
    \label{fig:ex_intro}
    \vspace{-.3cm}
\end{wrapfigure}

\section{Overview} \label{sec:overview}

\paragraph{Problem \& importance.}
We tackle the problem of molecular conformer generation (MCG), i.e. predicting the ensemble of low-energy 3D conformations of a small molecule solely based on the molecular graph (\cref{fig:ex_intro}). A single conformation is represented by the list of 3D coordinates for each atom in the respective molecule. In this work, we assume that the low-energy states are implicitly defined by the given dataset, i.e. our training data consist of molecular graphs and corresponding sets of energetically favorable 3D conformations. Low-energy structures are the most stable configurations and, thus, expected to be observed most often experimentally.

Dealing with molecules in their natural 3D structure is of great importance in areas such as cheminformatics or computational drug discovery because conformations determine biological, chemical, and physical properties \citep{guimaraes2012,schutt2018schnet,klicpera2019directional,axelrod2020molecular,schutt2021equivariant}  such as charge distribution, potential energy, docking poses~\citep{mcgann2011fred}, shape similarity~\citep{kumar2018advances}, pharmacophore searching~\citep{schwab2010conformations}, or descriptors for 3/4D QSAR~\citep{verma20103d}. For instance, in drug design it is crucial to  understand how a molecule binds to a specific target protein; this process heavily depends on the 3D structures of the  two components, both in terms of geometric (shape matching) and chemical (hydrophobic/hydrophilic) interactions  \citep{gainza2020deciphering,sverrisson2020fast}.

\paragraph{Motivation \& challenges of existing methods.} The main challenge in MCG comes from the enormous size of the 3D structure space consisting of bond lengths, bond angles, and torsion angles. It is known that the molecular graph imposes specific constraints on possible 3D conformations, e.g. bond length ranges depend on the respective bond types, while chiral centers dictate local spatial arrangement. However, the space of possible conformations grows exponentially with the graph size and number of rotatable bonds, thus hindering exhaustive brute force exploration even for relatively small molecules. Additionally, the number of plausibly-stable low-energy states is unknown a priori and can vary between one and several thousand conformations for a single molecule \citep{chan2021understanding}. Nevertheless, various facets of the curse of dimensionality have been favorably tackled by ML models in different contexts, and our goal is to build on the recent ML efforts for MCG~\citep{mansimov2019molecular,simm2020generative,lemm2021energy,xu2021learning}. 

Molecular conformations can be determined experimentally, but existing techniques are very expensive. As a consequence, predictive computational models have been developed over the past few decades, traditionally being categorized as either \textit{stochastic} or \textit{systematic} (rule-based)  methods~\citep{hawkins2017conformation}. Stochastic approaches have traditionally been based on molecular dynamics (MD) or Markov chain Monte Carlo (MCMC) techniques, potentially combined with genetic algorithms (GAs). They can do extensive explorations of the energy landscape and accurately sample equilibrium structures, but quickly become prohibitively slow for larger molecules \citep{shim2011computational,ballard2015exploiting,de2016role,hawkins2017conformation}, e.g. they require several CPU minutes for a single drug-like molecule. Moreover, stochastic methods have difficulties sampling diverse and representative conformers, prioritizing quantity over quality. On the other hand, rule-based systematic methods achieve state-of-the-art in commercial software~\citep{friedrich2017benchmarking} with \textsc{OMEGA}~\citep{hawkins2010conformer,hawkins2012conformer} being a popular example.  They usually process a single drug-like molecule under a second. They address the aforementioned challenges of stochastic methods by relying on carefully curated torsion templates (torsion rules), rule-based generators, and knowledge bases of rigid 3D fragments, which are assembled together and combined with subsequent stability score ranking. However, torsion angles are mostly varied independently (based on their fragments), without explicitly capturing their global interactions, which results in difficulties for larger and more flexible molecules. Furthermore, the curated fragments and rules are inadequate for more challenging inputs (e.g. transition states or open-shell molecules). 

Both types of methods can be combined with Distance Geometry (DG) techniques to generate the initial 3D conformation. First, the 3D atom distance matrix is generated based on a set of distance constraints or from a specialized model. Subsequently, the corresponding 3D atom  coordinates are learned to approximately match these predicted distances \citep{havel1983combinatorial,havel1983combinatorialbb,crippen1988distance,havel1998distance,lagorce2009dg,riniker2015better}. Indeed, modern stochastic algorithms are entirely based on DG methods \citep{riniker2015better}. The  inductive bias of rotational and translational invariance is guaranteed for DG, thus being appealing for ML models  \citep{simm2020generative,xu2021learning,pattanaik2020generating}. However, several drawbacks weaken this important direction: i) the distance matrix is overparameterized compared to the actual number of degrees of freedom, ii) it is difficult to enforce 3D Euclidean distance constraints as well as geometric graph constraints (e.g. on torsion angles or rings~\citep{riniker2015better}); iii) important aspects of molecular geometry are not explicitly modeled, e.g. torsion angles of rotatable bonds or chiral centers; iv) expensive force-field energy fine-tuning of the generated conformers is vital for a reasonable quality~\citep{xu2021learning,simm2020generative}; iv) the resulting multi-stage pipeline is prone to error accumulation as opposed to an end-to-end model.

Previous methods often rely on a force field (FF) energy function minimization to fine-tune the conformers. These are hand-designed energy models which use parameters estimated from experiment and/or computed from quantum mechanics (e.g. Universal Force Field \citep{rappe1992uff}, Merck Molecular Force Field \citep{halgren1996merck}). However, FFs are crude approximations of the true molecular potential energy surface \citep{kanal2018sobering}, limited in the interactions they can capture in biomolecules due to their strong assumptions~\citep{barman2015pushing}. In addition, FF energy optimization is relatively slow and increases error accumulation in a multi-pipeline method.

\paragraph{Relation to protein folding. } There has been impressive recent progress on modeling protein folding dynamics~\citep{ingraham2018learning,alquraishi2019end,noe2019boltzmann,senior2020improved}, where crystallized 3D structures are predicted solely from the amino-acid sequence using ML methods. However, molecules pose unique challenges, being highly branched graphs containing cycles, different types of bonds, and chirality information. This makes protein folding approaches not readily transferable to general molecular data.


\paragraph{Our key contributions \& model in a nutshell.} In this work, we investigate the question:

\begin{center}
\textit{Can we design fast ML generative models of high quality, representative, diverse, and generalizable low-energy 3D conformational ensembles from molecular graphs?}
\end{center}

To tackle this question, we propose \textsc{GeoMol} (shown in \cref{fig:model_overview}), exhibiting the following merits:
\begin{itemize}
\item It is end-to-end trainable, non-autoregressive, and does not rely on DG techniques (thus avoiding aforementioned drawbacks). More precisely, it outputs a minimal set of geometric quantities  (i.e. angles and distances) sufficient for full deterministic reconstruction of the 3D conformer. 

\item It models conformers in an SE(3)-invariant (translation/rotation) manner by design. This desirable inductive bias was previously either achieved using multi-step DG methods~\citep{simm2020generative} or not captured at all~\citep{mansimov2019molecular}. 

\item It explicitly models and predicts essential molecular geometry elements: torsion angles and local 3D structures (bond distances and bond angles adjacent to each atom). Together with the input molecular graph, these are used for k-hop distance computation at train time and full deterministic conformation assembly at test time. Crucially, we do not over-parameterize these predictions, i.e. a single torsion angle is computed per each non-terminal bond, irrespective of the number and permutation of the neighboring atoms at each end-point of the respective bond.

\item The above geometric elements (torsion angles, local structures) are SE(3)-invariant (by definition or usage) and we jointly predict them using MPNNs~\citep{gilmer2017neural} and self-attention networks. Thus, unlike~\citep{mansimov2019molecular}, we are not affected by MPNNs' pitfalls that obstruct direct predictions of 3D atom coordinates from node embeddings, e.g. symmetric or locally isomorphic nodes would always have identical MPNN embeddings~\citep{xu2018powerful,garg2020generalization} and, as a consequence,  would be inappropriately assigned identical 3D coordinates.
 
\item To promote diverse conformer ensembles with good coverage, we devise a tailored generative loss that does not use slow or difficult-to-optimize adversarial training techniques. Using optimal transport, \textsc{GeoMol} finds the best matching between generated and ground truth conformers based on their pairwise log-likelihood loss, requiring only minimization.

\item It explicitly and deterministically distinguishes reflected structures (enantiomers) by solving tetrahedral chiral centers using oriented volumes and local chiral descriptors, bypassing the need for iterative optimization usually done in DG approaches.

\item Empirically, we conduct experiments on two benchmarks: GEOM-QM9 (smaller molecules relevant to gas-phase chemistry) and GEOM-DRUGS (drug-like molecules) \citep{axelrod2020geom}. Our method often outperforms previous ML and two popular open-source  or commercial methods in different metrics. Moreover, we show competitive quality even without the frequently-used computationally-demanding fine-tuning FF strategies. 

\item \textsc{GeoMol} processes drug-like molecules in seconds or less, being orders of magnitude faster than popular baselines (e.g. ETKDG/RDKit\citep{riniker2015better}), without sacrificing quality.
\end{itemize}

%% file: method.tex
\begin{figure*}
    \vspace{-1cm}
    \centering
    \includegraphics[width=1.0\linewidth]{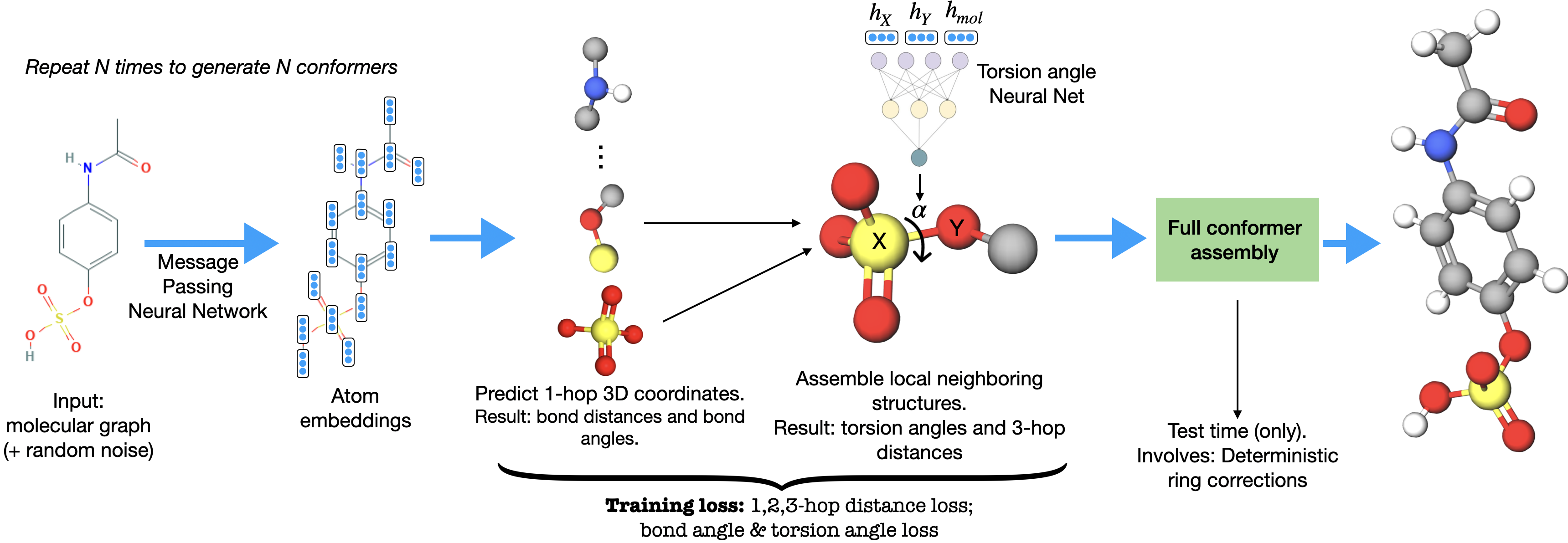}
    \vspace{-0.5cm}
    \caption{Overview of the \textsc{GeoMol} model, which is SE(3)-invariant by design. Given a  molecular graph, we first compute MPNNs atom embeddings. 
    Next, we predict the local 3D structures (LS) of each non-terminal atom in a permutation invariant way, explicitly solving chirality. Third, for each bond connecting non-terminal vertices, we assemble the two LS by predicting a single  torsion angle, avoiding overparameterization. Finally, the full conformer is assembled (only) at test time.}
    \label{fig:model_overview}
\end{figure*}

\section{Method}\label{sec:method}
\paragraph{Problem setup \& notations. } Our input is any molecular graph $G=(V,E)$ with node and edge features, $\x_v \in \R^{f}, \forall v \in V$ and $\e_{u,v} \in \R^{f'}, \forall (u,v) \in E$ representing atom types, formal charges, bond types, etc.  For each molecular graph G, we have a variable-size set of low-energy ground truth 3D conformers $\{\mathcal{C}^*_l\}_l$ that we  predict with a model $\{\mathcal{C}_k\}_k \myeq \zeta(G)$. A conformer is a map $\mathcal{C} : V \rightarrow \R^3$ from graph nodes to 3D coordinates, but a simplified notation is $\cc_v \in \R^3$ for $v \in V$. We use additional notations: $d(X,Y) = \|\cc_X - \cc_Y\|$ is the 3D distance between X and Y; $\angle XYZ$ is the counter-clockwise (CCW) angle $\angle \cc_X\cc_Y\cc_Z$; $\angle (XYZ, XYT)$ is the CCW dihedral angle of the 2D planes $\cc_X\cc_Y\cc_Z$ and $\cc_X\cc_Y\cc_T$ (formula is in \cref{apx:dihedral_formula}). We use the corresponding $\cc_v^*, d^*(X,Y), \angle^* XYZ, \angle^* (XYZ, XYT)$ when manipulating  a ground truth conformer.

Any conformer is defined up to a SE(3) transformation, i.e. any translation or rotation applied to the set $\{\cc_v \}_{v \in V}$. A classic conformer distance function that satisfies this constraint is root-mean-square deviation of atomic positions (RMSD), computed by the Kabsch alignment algorithm~\citep{kabsch1976solution}. 

\vspace{-0cm}

\subsection{\textsc{GeoMol} high-level overview} Our approach, shown in \cref{fig:model_overview}, comprises three steps. First, we predict the local 3D structure of each non-terminal atom, which we deem \textit{local structure} (LS), by combining self-attention layers and MPNNs with deterministic corrections for chiral centers. Bond distances and bond angles are computed from the predicted LS. Next, we assemble all neighboring pairs of LSs by predicting the torsion angles and aligning them. Importantly, since LSs are fixed, it suffices to only predict a single value for the dihedral angle of each bond. Towards this goal, we develop a canonical representation of torsion angles via a local coordinate system defined SE(3)-equivariantly  w.r.t. the full structure, which allows us to predict exactly the number of degrees of freedom. Finally, at test time, we assemble all predicted pairs of neighboring LSs to construct the full conformer, applying deterministic ring corrections. In order to generate diverse conformers, we append random Gaussian noise vectors to each initial node feature vector and use an optimal transport-based loss function for training. 

\subsection{Message passing neural networks (MPNNs)}  Given an input graph G, an MPNN~\citep{gilmer2017neural,battaglia2018relational,yang2019analyzing} computes node embeddings $\h_v \in \R^d, \forall v \in V$ using $T$ layers of iterative message passing:
\begin{equation}
\h_u^{(t+1)}=\psi\left(\h_u^{(t)}, \sum_{v\in \mathcal{N}_u} \phi(\h_v^{(t)}, \h_u^{(t)}, \e_{u,v}) \right) , \quad \text{where\ } \h_v^{(0)} \myeq concat[\x_v, \z_v],\ \z_v \sim \mathcal{N}(\mathbf{0},s \I_d)
\label{eq:mpnn}
\end{equation}
for each $t \in [0\ ..\ T-1]$, where $\mathcal{N}_u=\{v\in V |(u,v)\in E\}$, while $\psi$ and $\phi$ are generic functions, e.g. implemented using multilayer perceptrons (MLP) or  attention~\citep{velivckovic2017graph}. Final node embeddings are obtained by the embedding of the last layer: $\h_v \myeq \h_v^{(T)}, \forall v \in V$. Finally, we also compute a molecular embedding: $\h_{mol} \myeq MLP(\sum_{v \in V} \h_v)$.  We leave comparison with other MPNN variants for future work, e.g. \citet{kipf2016semi,velivckovic2017graph,hamilton2017inductive,xu2018powerful}.


\subsection{Local structure (1-hop) prediction model} \label{ssec:ls}
\begin{wrapfigure}{r}{0.35\textwidth}
\vspace{-.5cm}
  \begin{center}
    \includegraphics[width=0.15\textwidth]{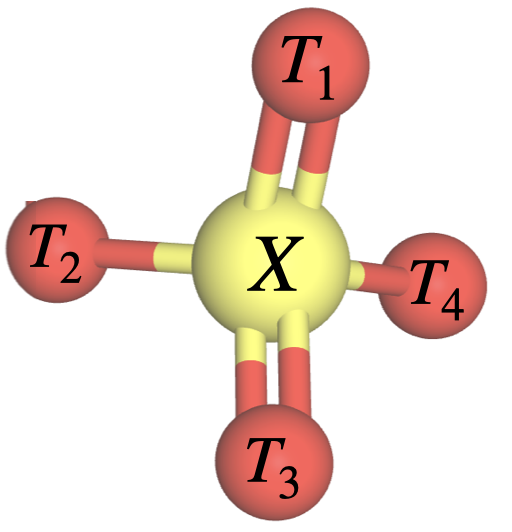}
    \vspace{-0.2cm}
  \end{center}
  \caption{For each non-terminal atom X, we predict the relative 3D position of each of its graph neighbors, $\{T_i\}_{i \in [1..n]}$, in a permutation equivariant way.}
    \label{fig:local_neigh}
\vspace{-0.5cm}
\end{wrapfigure}

Following notations in \cref{fig:local_neigh}, for each non-terminal graph vertex $X \in V$ having $n$ graph neighbors $\mathcal{N}_X = \{T_i\}_{i \in [1..n]}$, we predict its \textit{local 3D structure} (LS), i.e. the relative 3D positions of all $T_i$, when X is centered in the origin. The generic model is a function $f(\h_{T_1}, \ldots, \h_{T_n}; \h_X) = (\p_1, \ldots, \p_n) \in \R^{3 \times n} $ that, additionally, should satisfy permutation equivariance w.r.t. $T_i$'s, namely, the 3D position of each neighbor $T_i$ should not change regardless of the ordering of the $X$'s neighbors:
\begin{equation}
    f(\h_{T_{\pi(1)}}, \ldots, \h_{T_{\pi(n)}}; \h_X) = (\p_{\pi(1)}, \ldots, \p_{\pi(n)}), \forall \pi \in S_n
\end{equation}
Our choice is the encoder part of a transformer~\citep{vaswani2017attention}, without any positional encoding, thus satisfying permutation equivariance. This model takes as input the set $\{concat[\h_{T_i}, \h_{X}]; i  \in [1..n]\}$ \textit{in any order} and synchronously updates the $n$ embeddings based on several  transformer layers. The final layer projects the embeddings to 3 dimensions, resulting in a list $(\p_1, \ldots, \p_n) \in \R^{3 \times n}$ having the exact same node order as the input list.

\paragraph{Enforcing local consistency.} We desire the LS model $f()$ to be \textit{distance-consistent}, i.e. any bond distance $d(X,Y)$ is the same, no matter if it is computed from the LS of node X or of node Y. To achieve this, we use the above transformer just to compute bond directions (which will be aligned using a separate approach described in \cref{ssec:torsion}), while we obtain the bond distances with a separate symmetric model. Concretely, let the above transformer $f()$ predict $(\p_1, \ldots, \p_n) \in \R^{3 \times n}$, while the final local 3D coordinates are ${\p_i' \myeq \frac{\p_i}{\| \p_i\|} d_{GNN}(\h_X, \h_{T_i}), \forall i}$, where each bond distance is predicted with a symmetric model $d_{GNN}(\h_X, \h_{Y}) \myeq \text{softplus}(\psi(\h_X, \h_Y) + \psi(\h_Y, \h_X)), \forall (X,Y) \in E$, with the same shared $\psi$ (e.g. an MLP). For notation simplicity, we will just use $\p_i$ instead of  $\p_i'$.

\paragraph{What about SE(3) invariance?} The above model is not SE(3)-invariant \textit{per se}, but it is used as such. Namely, on one hand we compute SE(3)-invariant quantities: 1-hop distances $d(T_i,X)$, 2-hop distances $d(T_i, T_j)$, and bending angles $\angle T_iXT_j$. These will be compared to their ground-truth counterparts in the final loss, see \cref{ssec:loss}.  On the other hand, the LS of adjacent graph nodes are assembled together for computing torsion angles or for building the full conformer at test time. This process is explicitly defined to be SE(3)-invariant as described in \cref{ssec:torsion}.

\begin{wrapfigure}{r}{0.35\textwidth}
\vspace{-0.2cm}
  \begin{center}
    \includegraphics[width=0.33\textwidth]{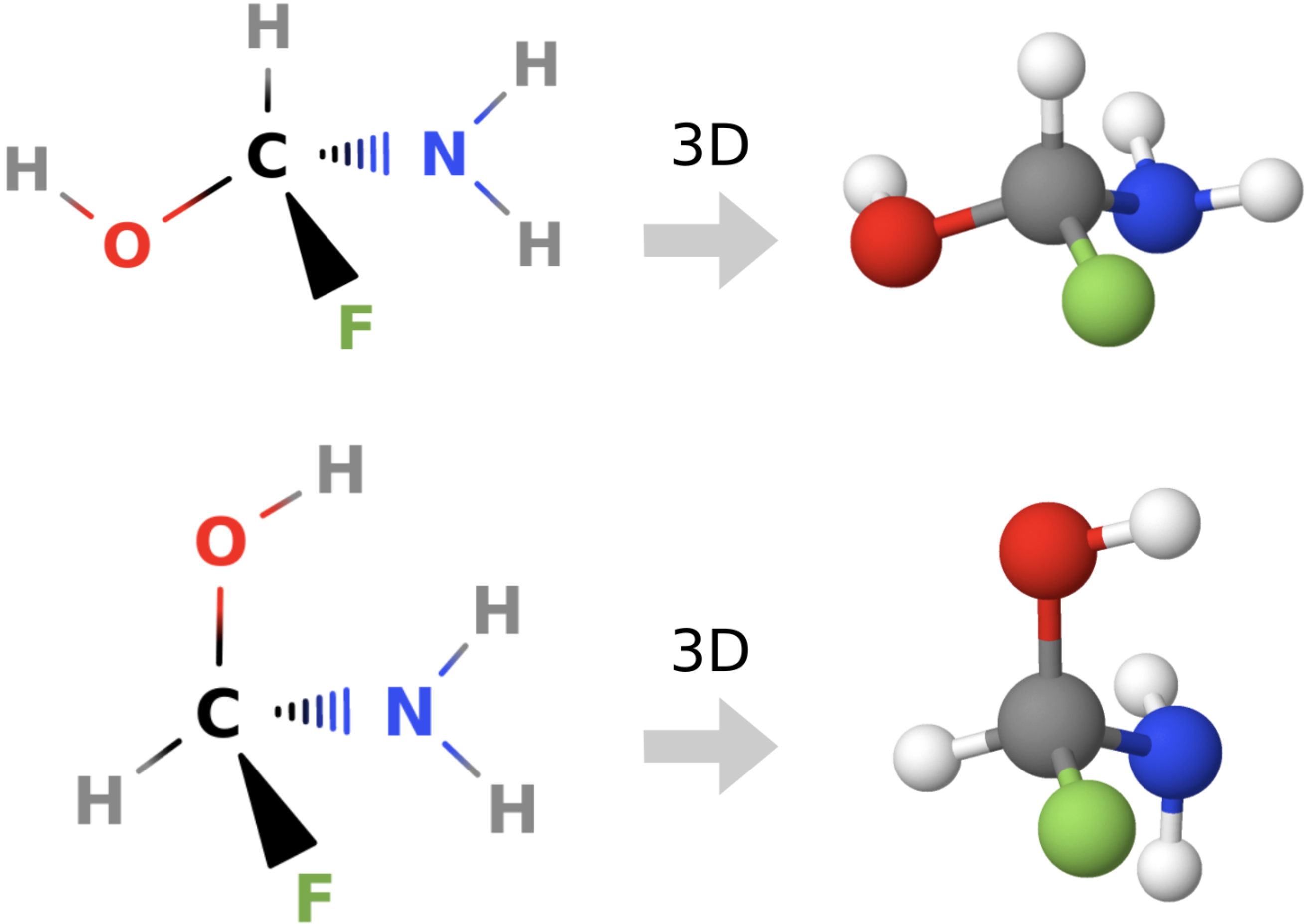}
    \vspace{-0.3cm}
  \end{center}
  \caption{Chirality: even if the two shown graphs are isomorphic, they have distinct 3D structures that can be distinguished by the order of the carbon center's neighbors.}
    \label{fig:chiral}
  \vspace{-1.3cm}
\end{wrapfigure}

\paragraph{Tetrahedral chiral corrections.} When embedding the local neighborhood of a node in 3D space, one has to carefully account for tetrahedral chiral centers (\cref{fig:chiral}). Tetrahedral chirality is a common form of stereochemistry which restricts the 3D location of neighboring substituents of a central atom with four distinct neighbors; molecules which differ by a single tetrahedral chiral center, i.e. enantiomers, are mirror images of each other. Chirality heavily impacts some properties of small
{
\parfillskip=0pt
\parskip=0pt
\par}
molecules--e.g. bioactivity. Existing MPNNs using only the molecular graph cannot distinguish chiral centers (\cref{fig:chiral}), but solutions exist~\citep{pattanaik2020message}. Mathematically, enantiomers can be differentiated based on the oriented volume around the chiral center. That is, given the ordered set of neighbor 3D coordinates around the center, namely $\p_1, \p_2, \p_3, \p_4 \in \R^3$, the sign of the volume of

the tetrahedron formed by the neighbors is
\begin{equation*}
    OV(\p_1, \p_2, \p_3, \p_4) \myeq sign \left(
    \begin{vmatrix}
    1 & 1 & 1 & 1\\
    x_1 & x_2 & x_3 & x_4\\
    y_1 & y_2 & y_3 & y_4\\
    z_1 & z_2 & z_3 & z_4\\
    \end{vmatrix} \right)
\end{equation*}
Enantiomeric structures always have opposite signs for the oriented volume \citep{crippen1988distance}. Since we generate local 3D structures directly, we can also use local 3D chiral descriptors to ensure the correct generation of tetrahedral chiral centers. RDKit internally keeps track of these local chiral labels, denoted by CW/CCW labels (detailed in e.g. \citet{pattanaik2020message}). Importantly, each local chiral label corresponds to a certain oriented volume (CW = +1 and CCW = -1). Thus, when generating an LS for a tetrahedral center, we calculate the oriented volume and check against the internal RDKit label. If it results in the incorrect oriented volume (i.e. the incorrect chiral center was generated), we simply reflect the structure by flipping against the z-axis. This ensures that all \textit{chiral centers are generated exactly}, and no iterative optimization is necessary as with traditional DG-based generators.


\subsection{Torsion angle representation and local structure (LS) assembly} \label{ssec:torsion}

\begin{wrapfigure}{r}{0.36\textwidth}
\vspace{-0.4cm}
  \begin{center}
    \includegraphics[width=0.4\textwidth]{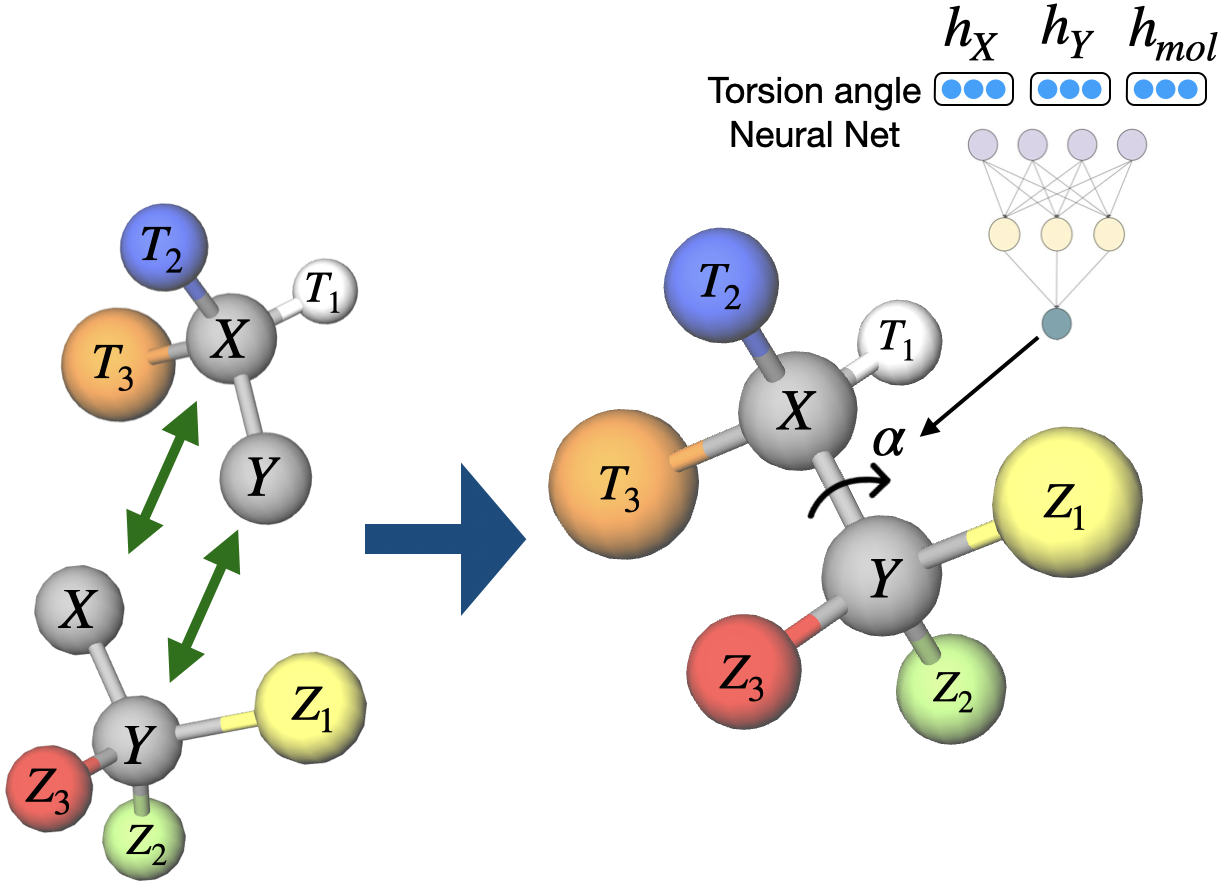}
    \vspace{-0.3cm}
  \end{center}
  \caption{Assembly of the local structures of bonded atoms X and Y based on the predicted torsion angle.}
    \label{fig:torsion}
  \vspace{-1.2cm}
\end{wrapfigure}

Once the LS of each atom/vertex is predicted, we assemble them in pairs corresponding to each non-terminal bond in the molecular graph. We describe this process for a bond connecting atoms X and Y, each having additional graph neighbors $\{T_i\}_{i \in [1..n]}$ and, resp., $\{Z_j\}_{j \in [1..m]}$. See \cref{fig:torsion}.

\paragraph{Torsion angle over-parameterization.} We first note that, for any assembled bond XY (\cref{fig:torsion}, right), and $\forall i,k \in [1..n], \forall j,l \in [1..m]$, the dihedral angles $\angle (XYT_i, XYT_k)$ and $\angle (XYZ_l, XYZ_j)$ are fully determined by the LS of nodes X and Y, respectively, so they do not depend on torsion angles. Next, observe that there is exactly one torsion angle for any bond XY, given unique indexing of the neighbors. This happens because of the constraint on $\angle (XYT_i, XYZ_j)$ and $\angle (XYT_k, XYZ_l)$:
\begin{equation}
    \angle (XYT_i, XYZ_j) = [\angle (XYT_k, XYZ_l) + \angle (XYT_i, XYT_k) + \angle (XYZ_l, XYZ_j)] (\mathrm{mod}\ 2\pi)
\end{equation}

Thus, in order to avoid unnecessary over-parameterization, we should predict a single torsion angle  $\alpha$. 

\paragraph{Torsion angle formulation. } However, it is still unclear at this point how to define this unique angle in a canonical way that is both: i) permutation invariant w.r.t. the nodes in the set $\{T_i\}_{i \in [1..n]}$ and, respectively, in the set $\{Z_j\}_{j \in [1..m]}$, and ii) SE(3)-invariant w.r.t. the full 3D conformer. 

Let $\Delta_{ij} \myeq \angle (XYT_i, XYZ_j)$ and $\s_{ij} \myeq \begin{bmatrix}
    \cos(\Delta_{ij}) \\
    \sin(\Delta_{ij})  
\end{bmatrix}$. Let $c_{ij} \in \R$ be real coefficients such that $\s \myeq \sum_{i,j} c_{ij} \s_{ij} \in \R^2$ is not the null vector. Then, we define the torsion angle as\footnote{We define atan2 slightly different than standard:  $atan2(r \cos(\alpha),r \sin(\alpha)) \myeq \alpha, \forall \alpha \in [0, 2\pi), r \in \R_+^*$.}:
${\alpha \myeq atan2(\frac{\s}{\|\s\|} )}$. It is easy to see that this formulation satisfies both invariances claimed above. We further state (and prove in \cref{proof:prop1}) that our proposed formulation gives a torsion angle  uniquely determined by all local angles $\angle (YXT_i, YXT_k)$, $\angle (YXZ_j, YXZ_l)$ and by the true underlying torsion angle:

\begin{proposition}
Given 3D coordinates of nodes $X,Y, T_i, Z_j$ and fixed weights $c_{ij} \in \R$ such that $\sum_{i,j} c_{ij} \s_{ij} \in \R^2$ is not the null vector, then $\alpha \myeq atan2(\frac{\s}{\|\s\|} )$ is unique, i.e. if we change the torsion angle of edge XY, then $\alpha$ will change. Formally, if we rotate the set of bonds $\{XT_i\}_i$ jointly around the line $XY$ with the same  angle $\gamma$, then $\alpha$ will be shifted with $\gamma$.
\label{prop1}
\end{proposition}

\paragraph{How to set $c_{ij}$? Breaking symmetries. } A simple solution is to choose $c_{ij} =1, \forall i,j$. However, in some important cases, local symmetries may result in $\s = 0$. For example, this happens if, for some $j$, we have $\Delta_{ij} = \frac{2i\pi}{n} + ct., \forall i \in [1..n]$. One  solution is to use different $c_{ij}$ to differentiate between the different subgraphs rooted at different $T_i$ (and similarly for $Z_j$). This is reminiscent of traditional group priorities used for distinguishing E/Z isomers. We devise a flexible solution: a differentiable real value function computed from the MPNN node embeddings as $c_{ij} = MLP(\h_{T_i} + \h_{Z_j}) \in \R$, with MLP being a network shared across all bonds and molecules. Note that we constrain $c_{ij} = c_{ji}$, thus guaranteeing that the same $\alpha$ is obtained if we swap X and Y (and their neighbors, respectively). 

\paragraph{Final LS assembly for a single bond.} We now describe the assembly process depicted in \cref{fig:torsion}. We first predict the LS of node X as in \cref{ssec:ls}, giving some $\p_X = \mathbf{0}, \p_Y, \p_{T_i}\in \R^3, \forall i \in [1..n]$, as well as the LS of node Y, giving $\q_Y = \mathbf{0}, \q_X, \q_{Z_j} \in \R^3, \forall j \in [1..m]$. By design, we have that $\|\q_X\| = \| \p_Y\|$. These two sets are currently not aligned. To achieve this, we first rotate the LS of X such that $\p_Y$ becomes $\begin{bmatrix}
          \| \p_Y\| &  0 & 0
\end{bmatrix}^{\top}$, while $\p_X$ remains $\mathbf{0}$. Next, we rotate and translate the LS of Y such that $\q_Y$ becomes $\p_Y$ and $\q_X$ becomes $\p_X = \mathbf{0}$. These two rotations have one degree of freedom each, which we set randomly. Exact formulas are in \cref{apx:random_assembly}. Thus, the bond XY is now matched, but the torsional rotation is still arbitrary/random. The  remaining step is to rotate the LS of X with an angle $\gamma$ such that all dihedrals $\angle (XYT_i, XYZ_j)$ match their true counterparts. This is done by applying to all vectors $\p_{T_i}$ the same rotation of type:
$\HH_\gamma := \begin{bmatrix}
          1 & 0 & 0 \\
          0 & \cos(\gamma) & -\sin(\gamma) \\
          0 &  \sin(\gamma) & \cos(\gamma)  
\end{bmatrix}$.

\textit{How to compute $\gamma$?} The current dihedrals $\Delta_{ij}^{cur} \myeq \angle^{cur} (XYT_i, XYZ_j)$ depend on the random torsional rotations from the previous assembly step. After applying the $\HH_\gamma$ rotation, we obtain the new dihedral angles: $[\Delta_{ij}^{cur} - \gamma] \text{\ mod\ } 2 \pi$ that should match the ground truth dihedral angles $\Delta_{ij}^* \myeq \angle^* (XYT_i, XYZ_j)$. This is equivalently written as $\s_{ij}^* = \A_{ij}^{cur}\s_{\gamma}$, where $\s_{\gamma} \myeq \begin{bmatrix}
    \cos(\gamma) \\
    \sin(\gamma)  
\end{bmatrix}$ and $\A_{ij}^{cur} \myeq \begin{bmatrix}
    \cos(\Delta_{ij}^{cur}) & \sin(\Delta_{ij}^{cur}) \\
    \sin(\Delta_{ij}^{cur}) & -\cos(\Delta_{ij}^{cur})  
\end{bmatrix}$. Let $\s^* \myeq \sum_{i,j} c_{ij}\s_{ij}^*$ and $\A^{cur} \myeq \sum_{i,j} c_{ij} \A_{ij}^{cur}$. The necessary condition for $\gamma$ becomes $\s_{\gamma} = \left(\A^{cur}\right)^{\top} \s^*$, which is also sufficient due to \cref{prop1}. This implies it is enough to predict \textit{only} the normalized $ \begin{bmatrix}
    \cos(\alpha) \\
    \sin(\alpha)  
\end{bmatrix} \myeq \frac{\s^*}{\|\s^*\|}$, from which $\gamma$ follows. We predict $\alpha$ using a function commutative in X and Y (i.e. swapping X and Y does not change $\alpha$):
\begin{equation}
\alpha = [\phi(\h_X, \h_Y, \h_{mol}) +  \phi(\h_Y, \h_X, \h_{mol}) ] \text{mod\ } 2\pi 
\label{eq:alpha}
\end{equation} where $\phi$ is a neural network (e.g. MLP).
Finally, 
$    \s_{\gamma} = \begin{bmatrix}
    \cos(\gamma) \\
    \sin(\gamma)  
\end{bmatrix} = \frac{1}{\|\left(\A^{cur}\right)^{\top} \s^* \|} \left(\A^{cur}\right)^{\top} \s^*$.

\subsection{An optimal transport (OT) loss function for diverse conformer generation}\label{ssec:loss}

\paragraph{Loss per single conformer. } Assume first that we predict a single conformer $\mathcal{C}$. Based on all LS and torsion angle predictions, we deterministically compute all 1/2/3-hop distances and bond/torsion angles. If the corresponding ground truth conformer $\mathcal{C}^*$ is known, we feed those quantities into a negative log-likelihood loss, denote by $\mathcal{L}(\mathcal{C}, \mathcal{C}^*)$ and detailed in \cref{apx:loss}. Similar to  \citet{senior2020improved}, we fit  distances using normal distributions and angles using von Mises distributions. This is a much faster approach compared to habitual RMSD losses that compare full conformers.

\begin{wrapfigure}{r}{0.3\textwidth}
\vspace{-.4cm}
  \begin{center}
    \includegraphics[width=0.3\textwidth]{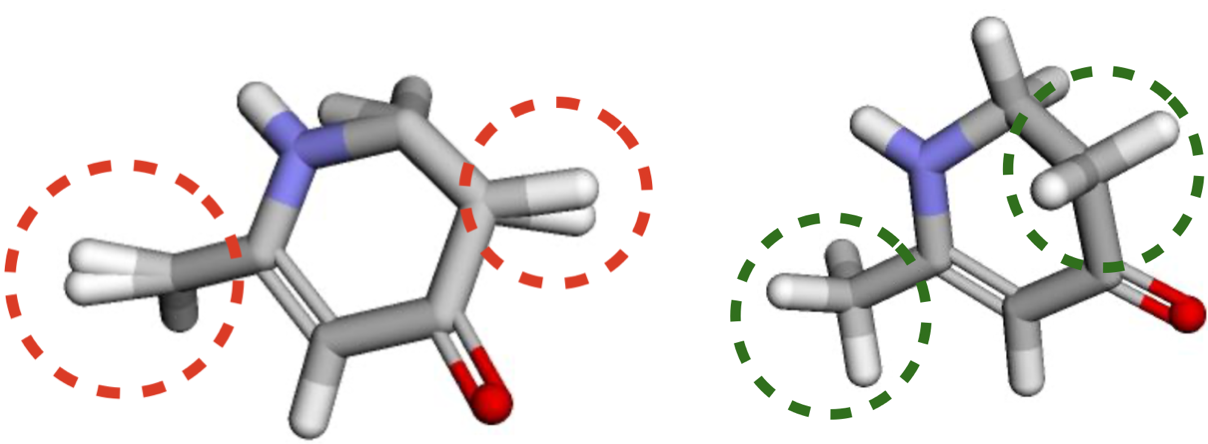}
    \vspace{-0.cm}
  \end{center}
  \caption{Before (left) and after (right) introducing a matching loss to distinguish symmetric graph nodes. Hydrogen predictions in both groups are visibly improved.}
    \label{fig:sym_corr}
  \vspace{0cm}
\end{wrapfigure}
\paragraph{Dealing with node symmetries. } Our current formulation has difficulties distinguishing pairs of symmetric graph nodes that are less than 3 hops away, e.g. hydrogen groups. We address this using a tailored matching loss detailed in \cref{apx:loss} and exemplified in \cref{fig:sym_corr}.

\paragraph{Total OT loss per ensemble of conformers. } 
In practice, our model generates a set of conformers $\{\mathcal{C}_k \}_{k \in [1..K]}$ that needs to match a variable sized set of low-energy ground truth conformers, $\{\mathcal{C}^*_l\}_{l \in [1..L]}$. However, we do not know a priori the number $L$ of true conformers or the matching between generated and true conformers. We also wish to avoid expensive and problematic adversarial training. Our solution is an OT-based, minimization-only, loss function: 

\begin{equation*}
    \mathcal{L}^{ensemble} \myeq \mathcal{W}_{\mathcal{L}(\cdot,\cdot)} (\{\mathcal{C}_k\}_k, \{\mathcal{C}^*_l\}_l) =  \min_{\T\in\mathcal{Q}_{K,L}} \sum_{k,l}T_{kl} \mathcal{L}(\mathcal{C}_k, \mathcal{C}^*_l)
    \label{eq:ot_loss}
\end{equation*}
where $\T$ is the \textbf{\textit{transport plan}} satisfying $\mathcal{Q}_{K,L} \myeq \{\T \in \R_+^{K\times L} : \T \mathbf{1}_L=\frac{1}{K}\mathbf{1}_K, \T^T \mathbf{1}_K=\frac{1}{L}\mathbf{1}_L \}$.
The minimization w.r.t. $\T$ is computed quickly using Earth Mover Distance and the POT library~\citep{flamary2017pot}.



\subsection{Full conformer assembly at test time}

Knowing all true LSs and torsion angles is, in theory, enough for a deterministic unique SE(3)-invariant reconstruction of the full conformer. However, in practice, these predictions might have small errors that accumulate, e.g. in rings. To mitigate this issue, we deterministically build the full conformer (only at test time) by first predicting a smoothed structure of (fused) rings separately, and then assembling the full conformer following any graph traversal order (any order gives the same conformer, so this procedure does not break the non-autoregressive behavior). We detail this step in \cref{apx:conf_assembly}.

%% file: exp.tex
\section{Experiments}\label{sec:exp}

We empirically evaluate \textsc{GeoMol} on the task of low-energy conformer ensemble generation for small and drug-like molecules. We largely follow the evaluation protocols of recent methods~\citep{simm2020generative,xu2021learning}, but also introduce new useful metrics.

\begin{table}
    \vspace{-0.4cm}
    \caption{Results on the \textbf{GEOM-DRUGS} dataset. All models are without FF fine-tuning. "R" and "P" denote Recall and Precision. Note: OMEGA is an established commercial (C) software.}
    \label{tab:drugs-res}
    \hspace{-0.4cm}
    \begin{tabular}{l||c|c||c|c||c|c||c|c }
    & \multicolumn{2}{c||}{\shortstack[c]{COV - R (\%) $\uparrow$}} & \multicolumn{2}{c||}{\shortstack[c]{AMR - R (\si{\angstrom}) $\downarrow$ }}  & \multicolumn{2}{c||}{\shortstack[c]{COV - P (\%) $\uparrow$}}  & \multicolumn{2}{c}{\shortstack[c]{AMR - P (\si{\angstrom}) $\downarrow$ }} \\
    \cline{2-9}
    Models & Mean & Median & Mean & Median & Mean & Median & Mean & Median \\
    \hline \hline
    $\mathrm{GraphDG}$ \textit{(ML)} & 10.37 & 0.00 & 1.950 & 1.933 & 3.98 & 0.00 & 2.420 & 2.420 \\ 
    $\mathrm{CGCF}$ \textit{(ML)} & 54.35 & 56.74 & 1.248 & 1.224 & 24.48 & 15.00 & 1.837 & 1.829 \\
    \hline
    $\mathrm{RDKit/ETKDG}$ & 68.78 & 76.04 & 1.042 & 0.982 & 71.06 & 88.24 & 1.036 & 0.943 \\ 
    $\mathrm{OMEGA}$ \textit{(C)} & 81.64 & 97.25 & 0.851 & \textbf{0.771} & 77.18 & \textbf{96.15} & 0.951 & \textbf{0.854} \\ 
     \hline \hline
    \textsc{GeoMol} ($s=9.5$) & \textbf{86.07} & \textbf{98.06} & \textbf{0.846} & 0.820 & 71.78 & 83.77 & 1.039 & 0.982 \\
    \textsc{GeoMol} ($s=5$) &  \shortstack[c]{82.43} & \shortstack[c]{95.10} & \shortstack[c]{0.862} & \shortstack[c]{0.837} & \shortstack[c]{\textbf{78.52}} & \shortstack[c]{94.40} & \shortstack[c]{\textbf{0.933}} & \shortstack[c]{\textbf{0.856}} 
      \end{tabular}
\end{table}
\begin{table}
    \caption{Results on the \textbf{GEOM-QM9} dataset. See caption of \cref{tab:drugs-res}.}
    \label{tab:qm9-res}
    \hspace{-0.4cm}
    \begin{tabular}{l||c|c||c|c||c|c||c|c }
     & \multicolumn{2}{c||}{\shortstack[c]{COV - R (\%) $\uparrow$}} & \multicolumn{2}{c||}{\shortstack[c]{AMR - R (\si{\angstrom}) $\downarrow$ }}  & \multicolumn{2}{c||}{\shortstack[c]{COV - P (\%) $\uparrow$}}  & \multicolumn{2}{c}{\shortstack[c]{AMR - P (\si{\angstrom}) $\downarrow$ }} \\
    \cline{2-9}
    Models & Mean & Median & Mean & Median & Mean & Median & Mean & Median \\
    \hline \hline
    $\mathrm{GraphDG}$ \textit{(ML)} & 74.66 & 100.00 & 0.373 & 0.337 & 63.03 & 77.60 & 0.450 & 0.404 \\ 
    $\mathrm{CGCF}$ \textit{(ML)} & 69.47 & 96.15 & 0.425 & 0.374 & 38.20 & 33.33 & 0.711 & 0.695 \\
    \hline
    $\mathrm{RDKit/ETKDG}$ & 85.13 & \textbf{100.00} & 0.235 & 0.199 & \textbf{86.80} & \textbf{100.00} & 0.232 & 0.205 \\ 
    $\mathrm{OMEGA}$ \textit{(C)} & 85.51 & \textbf{100.00} & \textbf{0.177} & \textbf{0.126} & 82.86 & \textbf{100.00} & \textbf{0.224} & \textbf{0.186} \\ 
    \hline \hline
    \textsc{GeoMol} ($s= 5$) & \shortstack[c]{\textbf{91.52}} & \shortstack[c]{\textbf{100.00}} & \shortstack[c]{0.225} & \shortstack[c]{0.193} & \shortstack[c]{\textbf{86.71}} & \shortstack[c]{\textbf{100.00}} & \shortstack[c]{0.270} & \shortstack[c]{0.241} \\
    \end{tabular}
\end{table}
\paragraph{Datasets \& splits. } We use two popular datasets: GEOM-QM9~\citep{ramakrishnan2014quantum} and  GEOM-DRUGS~\citep{axelrod2020geom}. Statistics and other details are in \cref{fig:dataset_stats} and in~\citet{mansimov2019molecular}. Datasets are preprocessed as described in \cref{apx:preprocessing}. We split them randomly based on molecules into train/validation/test  (80\%/10\%/10\%). At the end, for each dataset, we sample 1000 random test molecules as the final test set. Thus, the splits contain 106586/13323/1000 and 243473/30433/1000 molecules for GEOM-QM9 and GEOM-DRUGS, resp. 


\paragraph{Baselines. } We compare to established or recent baselines (discussed in \cref{sec:overview}). \textbf{ETKDG/RDKit} \citep{riniker2015better} is likely the most popular open-source software, a stochastic DG-based method developed in the RDKit package. \textbf{OMEGA}~\citep{hawkins2010conformer,hawkins2012conformer,friedrich2017benchmarking}, a rule-based method, is one of the most established commercial software, with more than a decade of continuous development. OMEGA and ETKDG are some of the fastest and best scaling existing approaches. Finally, we compare with the recent ML models of highest reported quality: \textbf{GraphDG}~\citep{simm2020generative} and \textbf{CGCF}~\citep{xu2021learning}.


\paragraph{Evaluation metrics. } We follow prior work~\citep{simm2020generative,xu2021learning} and use root-mean-square deviation of atomic positions (RMSD) to compare any two conformers. This is defined as the normalized Frobenius norm of the two corresponding matrices of 3D coordinates after being SE(3)-aligned a priori (using the Kabsch alignment algorithm~\citep{kabsch1976solution}). Next, we introduce four types of metrics to compare two conformer ensembles, generated by a method, $\{\mathcal{C}_k \}_{k \in [1..K]}$, and ground truth, $\{\mathcal{C}^*_l\}_{l \in [1..L]}$. These metrics follow the established classification metrics of Precision and Recall and are defined for a given threshold $\delta > 0$ as:
\vspace{-0.1cm}
\begin{equation}
\begin{split}
    & \text{COV - R (Recall)} \myeq \frac{1}{L} \left| \{l \in [1..L]: \exists k\in[1..K], RMSD(\mathcal{C}_k, \mathcal{C}^*_l) < \delta   \} \right| \\
    & \text{AMR - R (Recall)}  \myeq \frac{1}{L} \sum_{l \in [1..L]} \min_{k\in[1..K]} RMSD(\mathcal{C}_k, \mathcal{C}^*_l) \\
\end{split}
\label{eq:metrics}
\end{equation}
\vspace{-0cm}
where AMR is "Average Minimum RMSD", COV is "Coverage", and \textit{COV - P (Precision)} and \textit{AMR - P (Precision)} are defined as in \cref{eq:metrics}, but with the generated and ground truth conformer sets swapped. The recall metrics measure how many of the ground truth conformers are correctly predicted, while the precision metrics indicate how many generated structures are of high quality. Specifically, in terms of recall, COV measures the percentage of correct generated conformers from the ground truth set (where a correct conformer is defined as one within an RMSD threshold of the true conformer), while AMR measures the average RMSD of each generated conformer with its closest groun truth match. Depending on the application, either of the metrics might be of greater interest. We  follow~\citet{xu2021learning} and set $\delta=0.5 \si{\angstrom}$ for GEOM-QM9 and $\delta=1.25 \si{\angstrom}$ for GEOM-DRUGS.

\begin{figure}
  \hspace{-0.5cm}
    \includegraphics[width=1.03\textwidth]{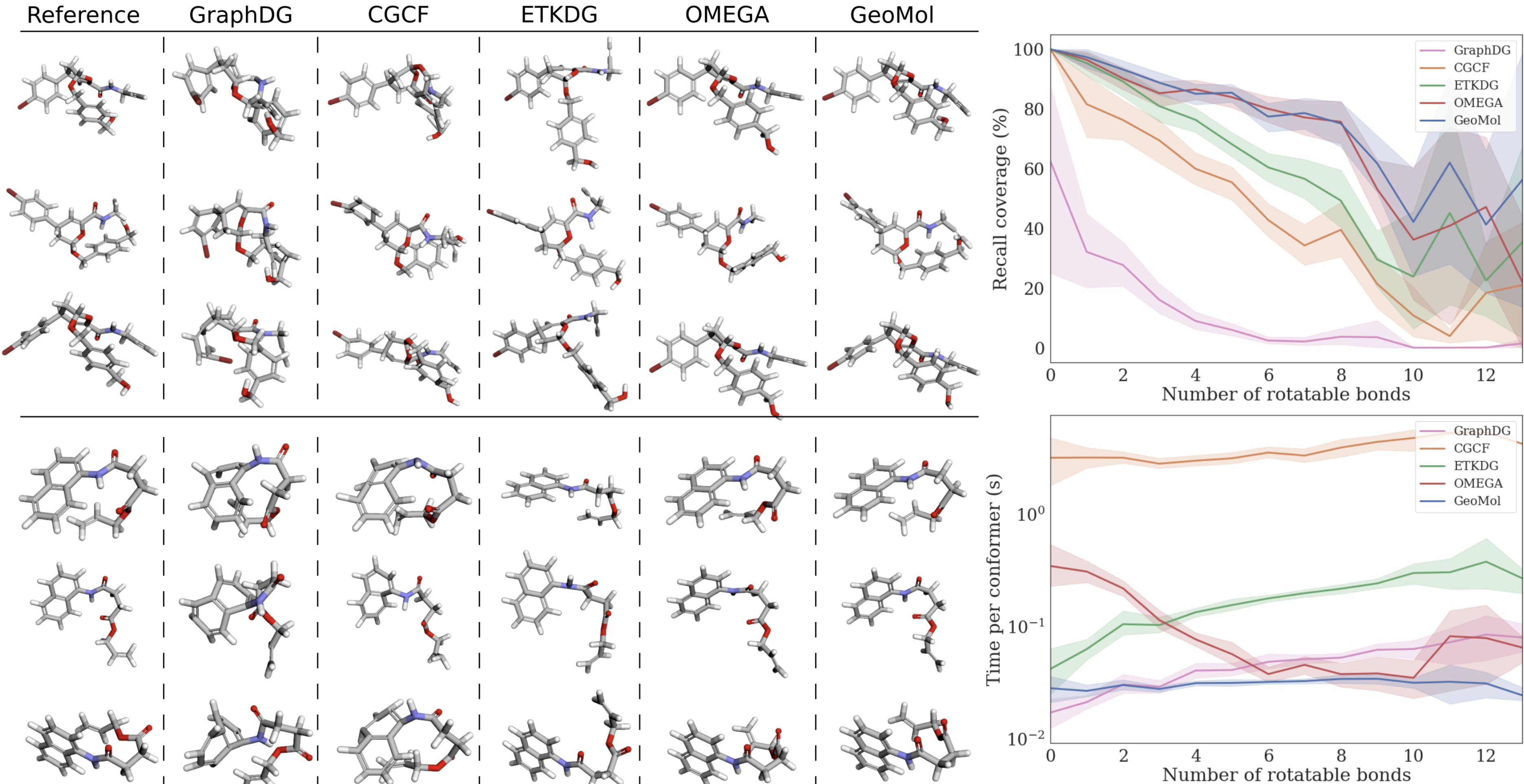}
  \caption{Left: Examples of generated structures. For every model, we show the best generated conformer, i.e. with the smallest RMSD to the shown ground truth. More examples are in \cref{apx:more_ex}. Right/top: Number of rotatable bonds per DRUGS test molecule versus COV Recall (95\% confidence intervals). Right/bottom: conformer generation times for each model.}
    \label{fig:examples}
\end{figure}

\paragraph{Training and test details.} For each input molecule having $K$ ground truth conformers, we generate exactly $2K$ conformers using any of the considered methods. For \textsc{GeoMol}, this is done by sampling different random noise vectors that are appended to node and edge features before the MPNN (\cref{eq:mpnn}). At train time, our model uses a standard deviation (std) $s$ (see \cref{eq:mpnn}) of 5 for both GEOM-QM9 and GEOM-DRUGS. At test time, \textsc{GeoMol} can use the same or different $s$ values, depending on the downstream application, i.e. higher $s$ results in more diverse conformers, while lower $s$ gives more quality (better precision). For OMEGA, it is not possible to specify a desired number of conformers. So, we tune the RMSD threshold (which decides how many conformers to keep) such that the total generated conformers by OMEGA are approximately $2K$. For GEOM-QM9, this corresponds to no RMSD cutoff (i.e. OMEGA generates all possible conformers), and for GEOM-DRUGS, this corresponds to a cutoff of 0.7\r{A} (meaning no two generated conformers will have a distance smaller than this cutoff). We discuss hyper-parameters and additional training details in \cref{apx:hyperparam}.

\paragraph{Results \& discussion. } Results are shown in \cref{tab:drugs-res} and \cref{tab:qm9-res}, and confidence intervals are in \cref{apx:confidence_intervals}. As noted above, \textsc{GeoMol} can be run with different noise std at test time, depending on which metric the user is interested in. Even though OMEGA is an established commercial software with more than a decade of continuous development, our model remarkably frequently outperforms it. Note that OMEGA fails to generate any conformers for 7\% of the QM9 test set (many of which include fused rings). Moreover, we also outperform the popular RDKit/ETKDG open-source model (except for AMR-P on QM9) and very recent ML models such as \textbf{GraphDG} and \textbf{CGCF}, sometimes by a large margin. 
For a qualitative insight, we show generated examples in \cref{fig:examples} and \cref{apx:more_ex}. 

Additionally, we show in \cref{fig:examples} how COV Recall results are affected by the increasing number of rotatable bonds in the test molecule. As expected, having more rotatable bonds makes the problem harder, and this affects all baselines, but \textsc{GeoMol} maintains a reasonable coverage even for more difficult molecules. We show additional results with energy-based relaxations in \cref{apx:ff_results}.



\paragraph{Running time. } Fig.~\ref{fig:examples} shows conformer generation test running times. Our model is the fastest method from the considered baselines, being much faster than \textbf{CGCF} or \textbf{ETKDG/RDKit}. Moreover, \textsc{GeoMol} scales favorably for molecules with increasing number of rotatable bonds. 

%% file: appendix.tex
\section{Proof of \cref{prop1}} \label{proof:prop1}

\begin{proof}
Let's assume we apply a random CCW torsion rotation of angle $\gamma \in (0, 2\pi)$ around bond XY to all bonds $XT_i$. We will prove that the resulting $\alpha$ will be shifted exactly by $\gamma$.  Let us denote the current dihedrals after the $\gamma$ rotation by $\Delta_{ij}^{cur} = \angle^{cur} (XYT_i, XYZ_j)  \text{\ mod\ } 2 \pi$. Also, denote by $\Delta_{ij} = \angle (XYT_i, XYZ_j)  \text{\ mod\ } 2 \pi$ the dihedral angles before rotation. We have the relation $\Delta_{ij} = [\Delta_{ij}^{cur} + \gamma] \text{\ mod\ } 2 \pi, \forall i,j$, or, in matrix form:
\begin{equation}
    \s_{ij} = \A_{\gamma}\s_{ij}^{cur}
\end{equation}
where 
$\A_{\gamma} := \begin{bmatrix}
    \cos(\gamma) & -\sin(\gamma) \\
    \sin(\gamma) & \cos(\gamma)  
\end{bmatrix}$, 
$\s_{ij} := \begin{bmatrix}
    \cos(\Delta_{ij}) \\
    \sin(\Delta_{ij})  
\end{bmatrix}$ and 
$\s_{ij}^{cur} := \begin{bmatrix}
    \cos(\Delta_{ij}^{cur}) \\
    \sin(\Delta_{ij}^{cur})  
\end{bmatrix}$. 

We know use $\s := \sum_{i,j} c_{ij} \s_{ij} \in \R^2$ and $\s^{cur} := \sum_{i,j} c_{ij} \s_{ij}^{cur} \in \R^2$. From the above relation, we obtain that $\frac{\s}{\|\s\|}  = \A_{\gamma}\frac{\s^{cur}}{\|\s^{cur}\|}$. Writing $\frac{\s}{\|\s\|} = \begin{bmatrix}
    \cos(\alpha) \\
    \sin(\alpha)  
\end{bmatrix}$ and $\frac{\s^{cur}}{\|\s^{cur}\|} = \begin{bmatrix}
    \cos(\alpha^{cur}) \\
    \sin(\alpha^{cur})  
\end{bmatrix}$ we obtain that $\alpha = \gamma + \alpha^{cur}$ which concludes our proof.
\end{proof}

\paragraph{A note on collinear 3D points. } In the main text, we always assumed that dihedral angles $\Delta_{ij} \myeq \angle (XYT_i, XYZ_j), \forall i \in [ 1.. n], \forall j \in [1 .. m]$  exist. However, any such angle will not be defined if X and Y are collinear with either $T_i$ or $Z_j$. In such cases, we identify those neighbor nodes that are predicted to be collinear with X and Y (solely based on their respective LS), and then remove them from the computation of $\A^{cur}$ and $\s_{\gamma}$. In the extreme case when these collinear nodes are the only neighbors, the torsion angle is not defined by our formulation. We leave this isolated case to be solved in future extensions of our work.

\section{Initial assembly of the LSs of the endpoints of a bond XY} \label{apx:random_assembly}

We detail here the formulae used in section \cref{ssec:torsion}. Let the (predicted) LS of node X be  $\p_X = \mathbf{0}, \p_Y, \p_{T_1}, \ldots, \p_{T_n} \in \R^3$, as in \cref{ssec:ls}. Similarly, let the LS of node Y be $\q_Y = \mathbf{0}, \q_X, \q_{Z_1}, \ldots, \q_{Z_m} \in \R^3$. By design of \cref{ssec:ls}, we have that $\|\q_X\| = \| \p_Y\|$. These sets of 3D points can be aligned by applying any SE(3) transformation to each of them. To achieve alignment, we first want to match the X and Y points.

For this, we first rotate the LS of X such that $\p_Y$ becomes $\begin{bmatrix}
          \| \p_Y\| \\
          0 \\
          0
\end{bmatrix}$ and $\p_X$ remains $\mathbf{0}$. Next, we rotate and translate the LS of Y such that $\q_Y$ becomes $\p_Y$ and $\q_X$ becomes $\mathbf{0}$. More concretely, let $\eta_X \in \R^3$ be any random unit-norm vector orthogonal to $\p_Y$. We find a rotation matrix $\HH_{X,Y} = \begin{bmatrix}
          - & \h_1^{\top} & -\\
          - &\h_2^{\top} & - \\
          - &\h_3^{\top} & -
\end{bmatrix} \in \R^{3 \times 3}$ such that $\HH_{X,Y}\p_Y = \begin{bmatrix}
          a \\
          0 \\
          0
\end{bmatrix}$, with $a > 0$, and  $\HH_{X,Y}\eta_X = \begin{bmatrix}
          b \\
          c \\
          0
\end{bmatrix}$, where $b,c \in \R$. This is solved as: 
$$\h_1 = \frac{\p_Y}{\|\p_Y \|}; \quad \h_3 =  \frac{\p_Y \times \eta_X}{\|\p_Y \times \eta_X \|}; \quad \h_2 = - (\h_1 \times \h_3)$$
We now rotate the LS of X by doing: $\p_{T_i}' \myeq \HH_{X,Y} \p_{T_i}, \forall i \in [1..n]$, $\p_{Y}' \myeq \HH_{X,Y}\p_{Y}$. We apply a similar procedure for the LS of Y, computing $\q_{Z_j}' \myeq \HH_{Y,X} \q_{Z_j}, \forall j \in [1..m]$ and $\q_{X}' := \HH_{Y,X} \q_{X}$. We now reflect the LS of Y w.r.t. the first and second coordinate by doing $\q_{Z_j}' \leftarrow \begin{bmatrix}
          -1 & 0 & 0 \\
          0 & -1 & 0\\
          0 & 0 & 1
\end{bmatrix} \q_{Z_j}' + \p_Y'$,
which centers X in the origin, i.e. $\q_{X}' = \textbf{0}$, and aligns the two vectors of Y, i.e. $\q_Y' = \p_Y'$.

\section{Dihedral angle formula} \label{apx:dihedral_formula}

The CCW dihedral angle between two intersecting half-planes ABC and ABD with common points A,B is computed as\footnote{See \url{https://en.wikipedia.org/wiki/Dihedral_angle}.} 
\begin{equation}
    \angle (ABC, ABD) = atan2 (\|\bb_2\|  \langle \bb_1, \bb_2 \times \bb_3 \rangle,  \langle \bb_1 \times \bb_2, \bb_2 \times \bb_3 \rangle)
    \label{eq:torsion_angle_generic}
\end{equation}
where $\bb_1 := \s_A - \s_C, \bb_2 := \s_B - \s_A, \bb_3:= \s_D - \s_B$.

\section{Details of the loss function} \label{apx:loss}

Assume, for a molecular graph $G=(V,E)$, that one predicted/generated conformer using one model is $\mathcal{C}$, while the corresponding ground truth conformer is $\mathcal{C}^*$. We use notations in \cref{sec:method}. $E$ is the set of edges in graph $G$. For every chain of 3 edges $ (u,v), (v,w), (w,y) \in E$, we define $\angle(uvwy) \myeq \angle(uvw, vwy)$.

Similar to AlphaFold \citep{senior2020improved}, we fit distances using normal distributions and angles using von Mises distributions. The resulting negative log-likelihood loss averages over same types of terms, being formally written as: 
\begin{equation}
\begin{split}
\mathcal{L}(\mathcal{C}, \mathcal{C^*}) \myeq &\ \xi_1 \cdot \frac{1}{\# \{(u,v) \in E\} }\sum_{\{(u,v) \in E\}} (d(u,v) - d^*(u,v))^2 \\
& + \xi_2 \cdot \frac{1}{\# \{u,v : \text{2-hops\ away}\} }\sum_{\{u,v : \text{2-hops\ away}\}} (d(u,v) - d^*(u,v))^2 \\
& + \xi_3 \cdot \frac{1}{\# \{u,v : \text{3-hops\ away}\} }\sum_{\{u,v : \text{3-hops\ away}\}} (d(u,v) - d^*(u,v))^2 \\
& - \xi_4 \cdot \frac{1}{\# (u,v) \in E, (v,w) \in E} \sum_{(u,v) \in E, (v,w) \in E} \cos\left(\angle uvw - \angle^* uvw\right) \\ 
& - \xi_5 \cdot \frac{1}{\# (u,v), (v,w), (w,y) \in E} \sum_{ (u,v), (v,w), (w,y) \in E} \cos\left(\angle(uvwy) - \angle^*(uvwy) \right) 
\end{split}
\label{eq:total_loss}
\end{equation}
where $\xi_1, \xi_2, \xi_3, \xi_4, \xi_5 > 0$ are hyperparameters tuned on the validation set, with $\xi_1, \xi_2, \xi_3$ representing inverse standard deviations of the respective normal distributions, while $\xi_4,\xi_5$ being the measures of concentration of the corresponding von Mises distributions.

\begin{wrapfigure}{r}{0.4\textwidth}
\vspace{-0.3cm}
  \begin{center}
    \includegraphics[width=0.25\textwidth]{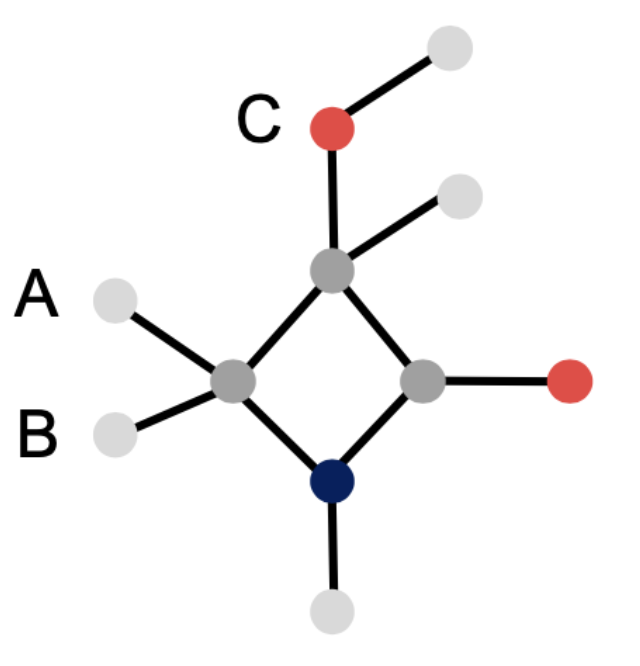}
  \end{center}
  \caption{Nodes A and B are symmetric in this example graph, but not when placed in the 3D space, as node A will be spatially closer to node C than node B. Such cases require a special treatment.}
    \label{fig:sym_ex}
    \vspace{-0.6cm}
\end{wrapfigure}

\paragraph{Dealing with node symmetries. A matching loss function.} So far, we haven't tackled the following difficulty: symmetric graph nodes that are less than 3 hops away are indistinguishable by MPNNs in general, and by our current model in particular, even if we append initial random noise feature vectors. However, these nodes have distinct distances and angles to other nodes in 3D space. Examples are hydrogen groups as in \cref{fig:sym_ex}. The difficulty in our case comes from the fact that the MPNN has lost the true matching between the graph nodes and the respective 3D points of the ground truth conformer. Our model is able to differentiate symmetric nodes because of the initial random noise vectors appended to each atom and feature vector, but a consistent matching with the ground truth conformer nodes is lost, and we seek to recover it. 

We here address this problem only for symmetric terminal nodes attached to the same common neighbor. We leave further extensions to more general symmetries for future work. Concretely, let $X_1, \ldots, X_T$ be identical atoms of degree 1, each being connected to the same common node $Y$ in the molecular graph. Let $\cc_1, \ldots , \cc_T \in \R^3$ be their predicted 3D coordinates in one generated  conformer $\mathcal{C} : V \rightarrow \R^3$. Let also $\cc_1^*, \ldots , \cc_T^* \in \R^3$ be their corresponding 3D coordinates of some ground truth conformer $\mathcal{C}^* : V \rightarrow \R^3$. These coordinates are then used for the computation of different loss terms in \cref{eq:total_loss}. 

We  propose a new loss function based on \cref{eq:total_loss} that replaces it:
\begin{equation}
    \mathcal{L}^{perm}(\mathcal{C}, \mathcal{C^*}) \myeq \min_{\pi \in S_T} \mathcal{L}(\mathcal{C}, \mathcal{C^*_\pi})
    \label{eq:perm_loss}
\end{equation}
where $\mathcal{C^*_\pi}  : V \rightarrow \R^3$ is defined as $\mathcal{C^*_\pi}(v) = \mathcal{C^*}(v), \forall v \in V \setminus \{X_1, \ldots, X_T\}$ and ${\mathcal{C^*_\pi}(X_i) = \mathcal{C^*}(X_{\pi(i)}), \forall i \in [1..T]}$. In a nutshell, this new loss function searches all possible permutations of the symmetric nodes $X_1, \ldots, X_T$ in order to find the "best matching" (based on the loss value) with the corresponding points in the ground truth conformer $\mathcal{C^*}$. The procedure above is applied for all groups of symmetric graph nodes that are of degree 1 and attached to a common node in the molecular graph.
The loss in \cref{eq:perm_loss} does not significantly affect the speed of our method (as we show in the experimental section) due to the fact that most groups of symmetric atoms contain at most 3 nodes. However, it positively affects the quality of the resulting conformers as exemplified in \cref{fig:sym_corr}.

\paragraph{OT Loss.} During training, the matrix $\T$ in \cref{eq:ot_loss} is computed for the forward pass using the Earth Mover Distance (EMD) function from the POT library~\citep{flamary2017pot} and kept fixed during backpropagation. The EMD computation cannot be parallelized in mini-batches in the current version of the library, but everything else is batch-parallelizable in our model (e.g. computation of $\mathcal{L}(\mathcal{C}, \mathcal{C^*})$).

\section{Details of the full conformer assembly procedure at test time} \label{apx:conf_assembly}

The training stage happens without assembling the full conformer. However, we desire to generate the full 3D conformer for test and inference time. This can be done by repeatedly applying the assembly operation described in \cref{ssec:torsion}. We assemble the 3D coordinates of the atoms one by one, following a fixed arbitrary graph traversal, say Breadth First Search (BFS). Our procedure described here is deterministic, and any graph traversal will lead to the same final conformer (up to an SE(3) transformation).

The key step is assembling two sets of 3D points: the set containing the LS of node X (and possibly other atom coordinates added in previous steps), denoted as $S_X := \{ \p_1, \ldots, \p_n\} \subset \R^3$, and the set containing the LS of node Y, denoted as $S_Y := \{ \q_1, \ldots, \q_m\} \subset \R^3$. Assume that X and Y are connected by a bond/edge. Also assume that X is connected to nodes $T_i$, while $Y$ is connected to nodes $Z_j$ as denoted in \cref{fig:torsion}. We make the clarification that  $\p_{T_i} \in S_X, \p_Y \in S_X, \p_X \in S_X$ and  $\q_{Z_j} \in S_Y, \q_Y \in S_Y, \q_X \in S_Y$ and we expect that $\|\p_X - \p_Y\| = \|\q_X - \q_Y\|$ by our design of this model.

To assemble the sets $S_X$ and $S_Y$, we follow the steps described in \cref{ssec:torsion} and \cref{apx:random_assembly}, but using only points in the LS of $X$ from the set $S_X$ together with only points of the LS of $Y$ from $S_Y$ to compute the torsion angle $\gamma$ of the bond $XY$. That is, the torsion angle is computed only based on the graph neighbors of nodes X and Y. However, all points in $S_X$ and $S_Y$ are used for all the updates of the 3D coordinates. In the end, we obtain assembled set of points that satisfy $\q_{Y} = \p_Y= \begin{bmatrix}
           d_{GNN}(\h_X, \h_Y)\\
           0 \\
           0
\end{bmatrix}, \q_{X} = \p_{X} = \textbf{0}$. Finally, we just merge the two sets $S_X$ and $S_Y$ into their union.

\begin{figure}
  \begin{center}
    \includegraphics[width=0.75\textwidth]{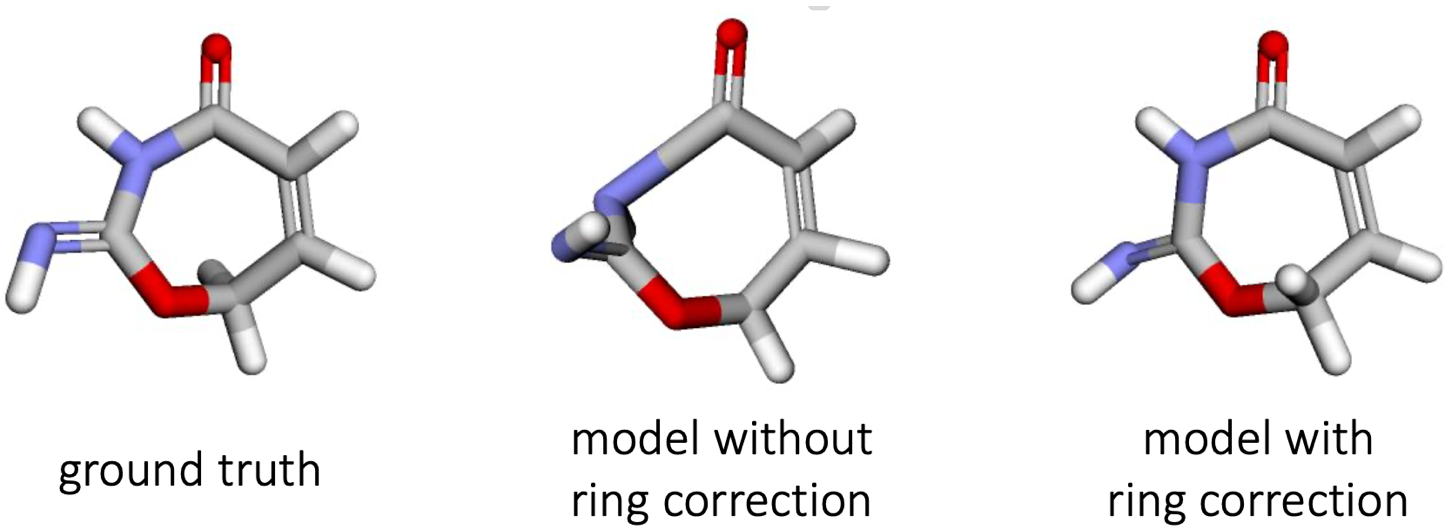}
  \end{center}
  \caption{Effect of two assembly techniques. Left: ground truth structure. Middle: our assembly procedure is applied on a random spanning tree of the molecular graph. Right: our described ring correction/smoothing algorithm, leading to a visibly improved structure. }
    \label{fig:ringcorr}
\end{figure}

\paragraph{Dealing with graph cycles. } The assembly operation might lead to inconsistencies when a node is reached the second time after a cycle traversal. We adopt a deterministic approach to tackle this issue. We first fix one BFS traversal of the graph and assemble the LS in the order given by this traversal. If the current node is not part of a cycle, then we can perform the assembling as described before. However, when we first encounter a node that is part of a cycle of nodes $X_1, X_2, \ldots , X_n$, we will jointly compute all the 3D coordinates of this cycle and attach the entire ring structure to the current partial conformer. Concretely, for every $i \in \{ 1, \ldots , n\}$, we assemble the LS of nodes $i, i+1 , \ldots n , 1, 2, \ldots i-2$ resulting in the 3D points $\x_{i-1}, \x_i \x_{i+1}, \ldots , \x_n, \x_1, \x_2 , \ldots , \x_{i-2}, \x_{i-1}'$, where node $i-1$ has two 3D points $\x_{i-1}$ and $\x_{i-1}'$ computed based on the LS of nodes $i$ and $i-2$ respectively. Next, we average the two 3D points of node $i-1$ into a single vector denoted as $\x_{i-1} \in \R^3$. We denote the resulting list of 3D points as $S_i = \{\x_{i-1}, \x_i \x_{i+1}, \ldots , \x_n, \x_1, \x_2 , \ldots , \x_{i-2}\}$. So, we have obtained $n$ lists of 3D points, $S_i, \forall i \in [1..n]$, that each represents the 3D conformer of the same cycle when the assembling is done using all nodes except one, in turn. We align all $S_i$ together using Kabsch algorithm~\citep{kabsch1976solution}~\footnote{\url{https://en.wikipedia.org/wiki/Kabsch_algorithm}}, followed by averaging all vectors of the same cycle node, for each node in  the cycle, thus obtaining a final smoothed set of 3D coordinates for this respective cycle. We call this procedure "ring correction" or "ring smoothing". We additionally note that, when aligning two sets $S_i$ and $S_j$, we also have to align the non-cycle points previously assembled. Finally, after all 3D vectors of a cycle are obtained, we continue to assemble in the order given by the BFS traversal. We show an example of the effect of this procedure in~\cref{fig:ringcorr}.

\section{Datasets statistics} \label{apx:datasets}

\begin{figure}[H]
    \hspace{-0.7cm}
    \includegraphics[width=1.05\linewidth]{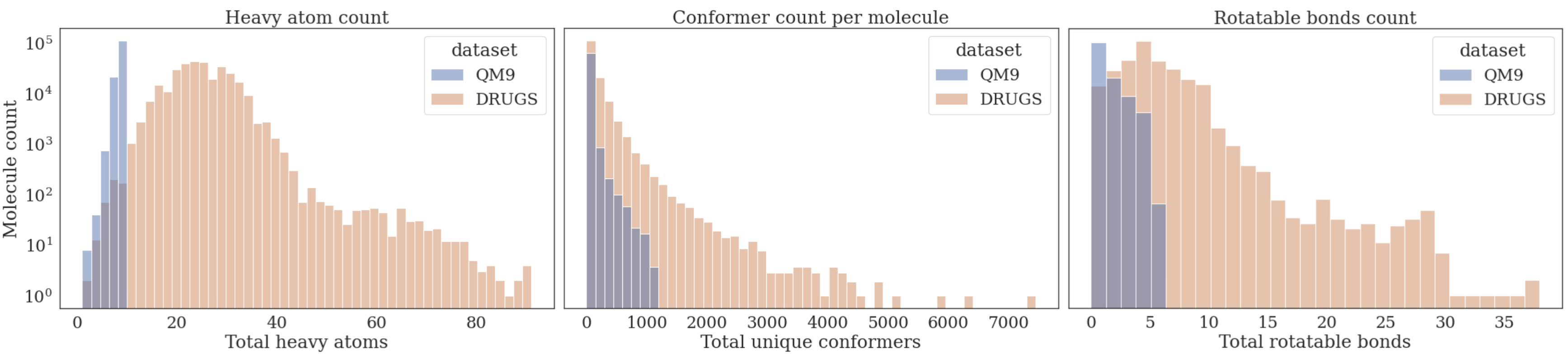}
    \caption{Datasets statistics.}
    \label{fig:dataset_stats}
\end{figure}
We here show several dataset statistics in \cref{fig:dataset_stats}, but point the interested reader to other resources describing these datasets, e.g. \citet{mansimov2019molecular,axelrod2020geom,ramakrishnan2014quantum}. 

\paragraph{Datasets generation details.} The GEOM dataset is generated with semi-empirical tight-binding DFT (GFN2-xTB) with the CREST software~\cite{grimme2019exploration}. However, such semi-empirical calculations can be relatively slow in practice--espeically for larger molecules--which is one of the main motivations of ML models for conformer ensemble prediction.

\section{Data preprocessing} \label{apx:preprocessing}

We featurize each .pickle file in the datasets, which correspond to a single molecule and its respective conformers, to PyTorch Geometric data objects. First, we ensure the provided SMILES strings can be processed by RDKit, and we discard the few molecules that fail to meet this criteria. We additionally discard molecules with disconnected fragments (i.e. contains a “.” in the SMILES) and without any dihedral angles (i.e. fails to match a “$[*]\sim[*]\sim[*]\sim[*]$” SMARTS pattern). Finally, some conformers in the original dataset may have reacted during the original data generation process. We filter these conformers by inferring the SMILES from the 3D structure, canonicalizing the SMILES with RDKit, and checking this SMILES with the SMILES reported by the dataset. Any conformers who fail this check are discarded as reacted conformers. We detail the atom and bond features used for the final GEOM-DRUGS model in Table \ref{tab:feats}. Note that the GEOM-QM9 model uses the same features but restricts the possible atom identities to H, C, N, O, and F.

\begin{table}[H]
    \centering
    \caption{Atom and bond features}\label{tab:feats}
    \begin{tabular}{cccc}
        \toprule
        \multicolumn{4}{c}{\multirow{2}{*}{Atom features}} \\ \\
        Indices & Description & Options & Type \\
        \midrule
        \shortstack[c]{0-34}    & \shortstack[c]{atom identity}                 & \shortstack[c]{H, Li, B, C, N, O, F, Na, Mg, Al, Si, P, S,\\
                                                                Cl, K, Ca, V, Cr, Mn, Cu, Zn, Ga, Ge,\\
                                                                As, Se, Br, Ag, In, Sb, I, Gd, Pt, Au, Hg, Bi}         & one-hot  \\
        35      & atomic number                         & $\mathbb{Z}_{>0}$                                 & value  \\
        36      & aromaticity                           & true, false                                       & one-hot  \\
        37-44   & degree                                & 0, 1, 2, 3, 4, 5, 6, other                        & one-hot  \\
        45-50   & hybridization                         & $sp$, $sp^2$, $sp^3$, $sp^3d$, $sp^3d^2$, other   & one-hot  \\
        51-58   & implicit valence                      & 0, 1, 2, 3, 4, 5, 6, other                        & one-hot  \\
        59-62   & formal charge                         & -1, 0, 1, other                                   & one-hot  \\
        63-69   & presence of atom in ring of size x    & 3, 4, 5, 6, 7, 8, other                           & k-hot  \\
        70-74   & number of rings atom is in            & 0, 1, 2, 3, other                                 & one-hot  \\
        \midrule
        \multicolumn{4}{c}{\multirow{2}{*}{Bond features}} \\ \\
        Indices & Description & Options & Type \\
        \midrule
        0-3  & bond type    & single, double, triple, aromatic  & one-hot  \\
        \bottomrule
    \end{tabular}
\end{table}

\section{Additional training details and hyperparameters} \label{apx:hyperparam} 

We run each model for a fixed 250 epochs with early stopping and keep the model with the best validation performance. This strategy determines the final hyperparameters chosen for the model, shown in \cref{tab:hyperparams}. We additionally tune the standard deviation $s$ on a sample of the validation set to maximize recall coverage or precision coverage (depending on the user's interest), and we show the corresponding test scores in our tables. To adjust the learning rate during training, we use a plateau scheduler which automatically decays the learning rate when improvement has stalled, using a decay factor of 0.7 and a patience of 5 epochs. All models use the same seed to initialize network weights, ensuring consistency between runs. During training time, we randomly sample a fixed number of ground truth conformers from each molecule. This fixed number is a hyperparameter in our optimization, which we set to 10 for GEOM-QM9 and 20 for GEOM-DRUGS. This value is not relevant during test time, as the model simply generates the number of conformers requested by the user.

\begin{table}
    \centering
    \caption{Hyerparameter choices}\label{tab:hyperparams}
    \begin{tabular}{lcc}
        \toprule
        Hyperparameter & Final choice (QM9/DRUGS) \\
        \midrule
        True conformers sampled           & 10/20 \\
        Model conformers generated        & 10/20 \\
        Model dimension                   & 25    \\
        Random vector dimension           & 10    \\
        Random vector standard deviation $s$  & 5     \\
        MPNN 1 depth                      & 3     \\
        MPNN 1 number of layers           & 2     \\
        MPNN 2 depth                      & 3     \\
        MPNN 2 number of layers           & 2     \\
        LS self-attention encoder heads   & 2     \\
        LS coordinate MLP layers          & 2     \\
        LS distance MLP layers            & 1     \\
        $h_{mol}$ MLP layers              & 1     \\
        $\alpha$ MLP layers               & 2     \\
        $c_{ij}$ MLP layers               & 1     \\
        Batch size                        & 16    \\
        Learning rate                     & 1e-3  \\
        \bottomrule
    \end{tabular}
\end{table}

\section{Confidence intervals} \label{apx:confidence_intervals}
We report in \cref{tab:confidence} confidence intervals for \textsc{GeoMol} on both datasets during test time. The same trained model is run for 3 times for a given test molecule with K ground truth conformers, where one run means generating $2K$ conformers.
\begin{table}
    \caption{Confidence intervals for \textsc{GeoMol} with the default noise $s=5$.}
    \label{tab:confidence}
    \hspace{-0.4cm}
    \begin{tabular}{c||c|c||c|c||c|c||c|c }
     & \multicolumn{2}{c||}{\shortstack[c]{COV - R (\%) $\uparrow$}} & \multicolumn{2}{c||}{\shortstack[c]{AMR - R (\si{\angstrom}) $\downarrow$ }}  & \multicolumn{2}{c||}{\shortstack[c]{COV - P (\%) $\uparrow$}}  & \multicolumn{2}{c}{\shortstack[c]{AMR - P (\si{\angstrom}) $\downarrow$ }} \\
    \cline{2-9}
    Models & Mean & Median & Mean & Median & Mean & Median & Mean & Median \\
    \hline \hline
    \shortstack[c]{\textsc{GeoMol}\\DRUGS} &  \shortstack[c]{82.43\\ $\pm$ 0.26} & \shortstack[c]{95.10 \\ $\pm$ 0.25} & \shortstack[c]{0.8626 \\ $\pm$ 0.0020} & \shortstack[c]{0.8374 \\ $\pm$ 0.0037} & \shortstack[c]{78.52 \\ $\pm$ 0.02} & \shortstack[c]{94.40 \\ $\pm$ 0.20} & \shortstack[c]{0.9336 \\ $\pm$ 0.0015} & \shortstack[c]{0.8567 \\ $\pm$ 0.0030} \\
    \hline \hline
    \shortstack[c]{\textsc{GeoMol}\\QM9} & \shortstack[c]{91.52\\ $\pm$ 0.36} & \shortstack[c]{100.00 \\ $\pm$ 0.00} & \shortstack[c]{0.2254 \\ $\pm$ 0.0012} & \shortstack[c]{0.1938 \\ $\pm$ 0.0024} & \shortstack[c]{86.71 \\ $\pm$ 0.29} & \shortstack[c]{100.00 \\ $\pm$ 0.00} & \shortstack[c]{0.2702 \\ $\pm$ 0.0007} & \shortstack[c]{0.2413 \\ $\pm$ 0.0030} \\
    \end{tabular}
\end{table}

\section{Results with force field optimization} \label{apx:ff_results}

We also show results in \cref{tab:drugs_ff-res} and \cref{tab:qm9_ff-res} for all models after additional fine-tuning with FF energy minimization. We used MMFF94s Merck Molecular Force Field \citep{halgren1996merck,halgren1999mmff}). Note that the best \textsc{GeoMol} models here use $s=1$ during training rather than $s=5$.

\begin{table*}
    \caption{Results on the \textbf{GEOM-DRUGS} dataset with FF fine-tuning. }
    \label{tab:drugs_ff-res}
    \hspace{-0.4cm}
    \begin{tabular}{l||c|c||c|c||c|c||c|c }
     & \multicolumn{2}{c||}{\shortstack[c]{COV - R (\%) $\uparrow$}} & \multicolumn{2}{c||}{\shortstack[c]{AMR - R (\si{\angstrom}) $\downarrow$ }}  & \multicolumn{2}{c||}{\shortstack[c]{COV - P (\%) $\uparrow$}}  & \multicolumn{2}{c}{\shortstack[c]{AMR - P (\si{\angstrom}) $\downarrow$ }} \\
    \cline{2-9}
    Models & Mean & Median & Mean & Median & Mean & Median & Mean & Median \\
    \hline \hline
    $\mathrm{GraphDG}$ \textit{(ML)} & 85.33 & 100.00 & 0.859 & 0.831 & 62.65 & 69.45 & 1.162 & 1.121 \\ 
    $\mathrm{CGCF}$ \textit{(ML)} & \textbf{91.25} & \textbf{100.00} & 0.723 & 0.702 & 63.86 & 70.43 & 1.096 & 1.055 \\ 
    \hline
    $\mathrm{RDKit/ETKDG}$ & 77.22 & 87.62 & 0.886 & 0.837 & 75.98 & 93.86 & 0.911 & 0.803 \\ 
    $\mathrm{OMEGA}$ \textit{(C)} & 80.82 & 94.74 & 0.840 & 0.755 & 81.79 & \textbf{100.00} & 0.804 & 0.684 \\
     \hline \hline
    \textsc{GeoMol} ($s=0.25$) & 80.73 & 92.01 & 0.863 & 0.809 & \textbf{86.12} & \textbf{100.00} & \textbf{0.751} & \textbf{0.616} \\
    \textsc{GeoMol} ($s=3.00$) & \textbf{91.34} & \textbf{100.00} & \textbf{0.683} & \textbf{0.663} & 79.64 & 92.46 & 0.841 & 0.756 \\
      \end{tabular}
\end{table*}
\begin{table*}
    \caption{Results on the \textbf{GEOM-QM9} dataset with FF fine-tuning. }
    \label{tab:qm9_ff-res}
    \hspace{-0.4cm}
    \begin{tabular}{l||c|c||c|c||c|c||c|c }
     & \multicolumn{2}{c||}{\shortstack[c]{COV - R (\%) $\uparrow$}} & \multicolumn{2}{c||}{\shortstack[c]{AMR - R (\si{\angstrom}) $\downarrow$ }}  & \multicolumn{2}{c||}{\shortstack[c]{COV - P (\%) $\uparrow$}}  & \multicolumn{2}{c}{\shortstack[c]{AMR - P (\si{\angstrom}) $\downarrow$ }} \\
    \cline{2-9}
    Models & Mean & Median & Mean & Median & Mean & Median & Mean & Median \\
    \hline \hline
    $\mathrm{GraphDG}$ \textit{(ML)} & 88.70 & 100.00 & 0.210 & 0.165 & 90.14 & \textbf{100.00} & 0.185 & 0.129 \\ 
    $\mathrm{CGCF}$ \textit{(ML)} & 75.45 & 100.00 & 0.313 & 0.246 & 50.29 & 50.00 & 0.518 & 0.520 \\  
    \hline
    $\mathrm{RDKit/ETKDG}$ & 83.48 & 100.00 & 0.219 & 0.172 & 89.78 & \textbf{100.00} & \textbf{0.160} & \textbf{0.116} \\ 
    $\mathrm{OMEGA}$ \textit{(C)}& 85.73 & 100.00 & \textbf{0.177} & \textbf{0.126} & 83.08 & \textbf{100.00} & 0.224 & 0.186 \\ 
    \hline \hline
    \textsc{GeoMol} ($s=5.00$) & \textbf{89.37} & 100.00 & 0.201 & 0.157 & \textbf{91.92} & \textbf{100.00} & 0.173 & 0.124 \\
    \end{tabular}
\end{table*}

\section{Compute details} \label{apx:compute_details}

We train \textsc{GeoMol} using a single Nvidia Volta V100 GPU and eight Intel Xeon Gold 6248 CPUs. The exact number of hours depends on the cluster load and on early stopping, but training \textsc{GeoMol} for the full 250 epochs takes approximately 1.8 days for the QM9 dataset and 6.1 days for the DRUGS dataset.

\section{Additional coverage results}
\begin{figure}[H]
    \includegraphics[width=1.0\linewidth]{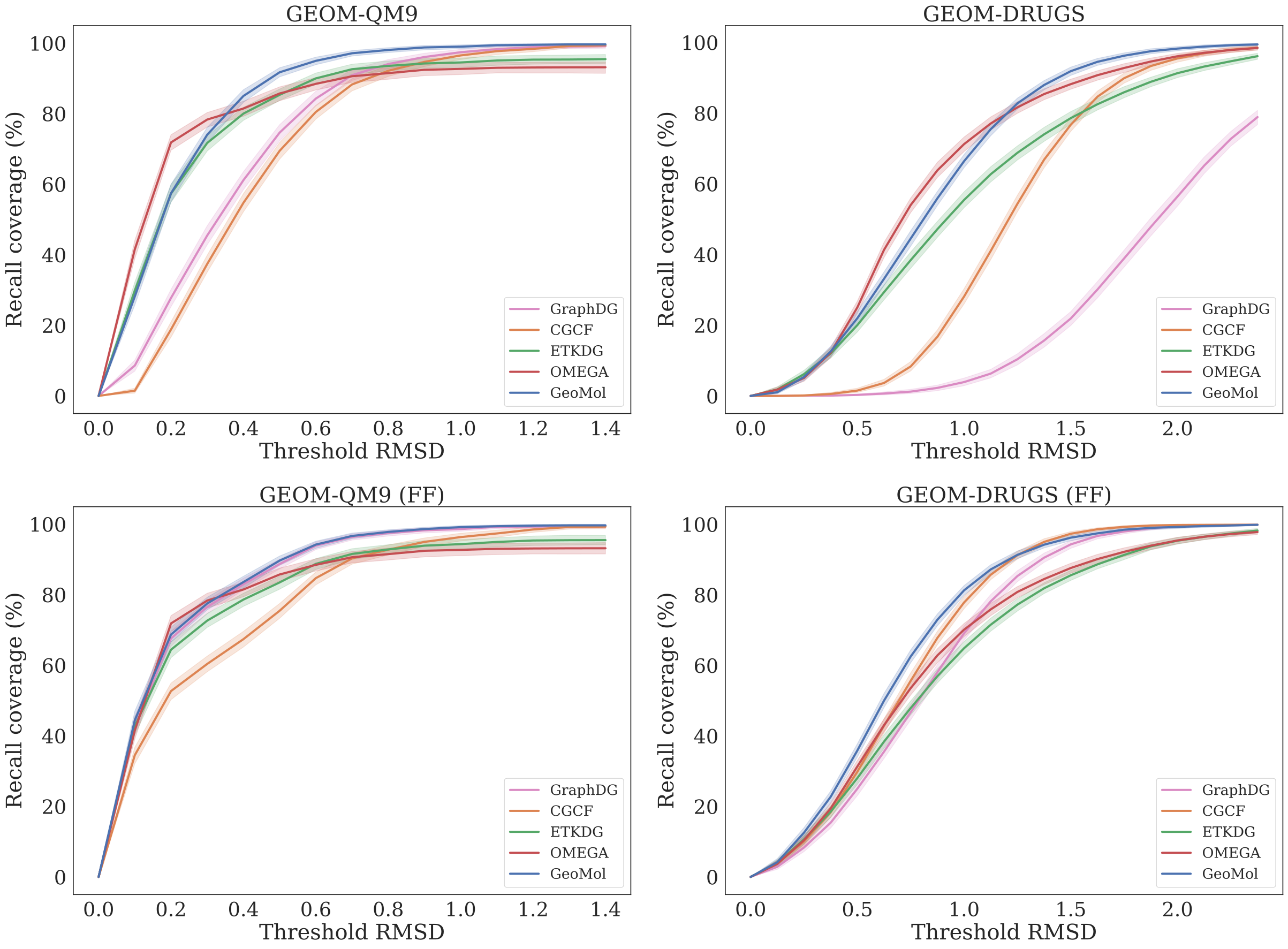}
    \caption{Recall coverage plotted against varying RMSD thresholds.}
    \label{fig:add_cov_res}
\end{figure}

We additionally show how the Recall Coverage varies with the threshold $\delta$ in \cref{fig:add_cov_res}. Larger thresholds allow small errors, but penalize large errors, while smaller thresholds penalize both small and large errors.

\section{Additional examples} \label{apx:more_ex}
\begin{figure}
    \includegraphics[width=1\linewidth]{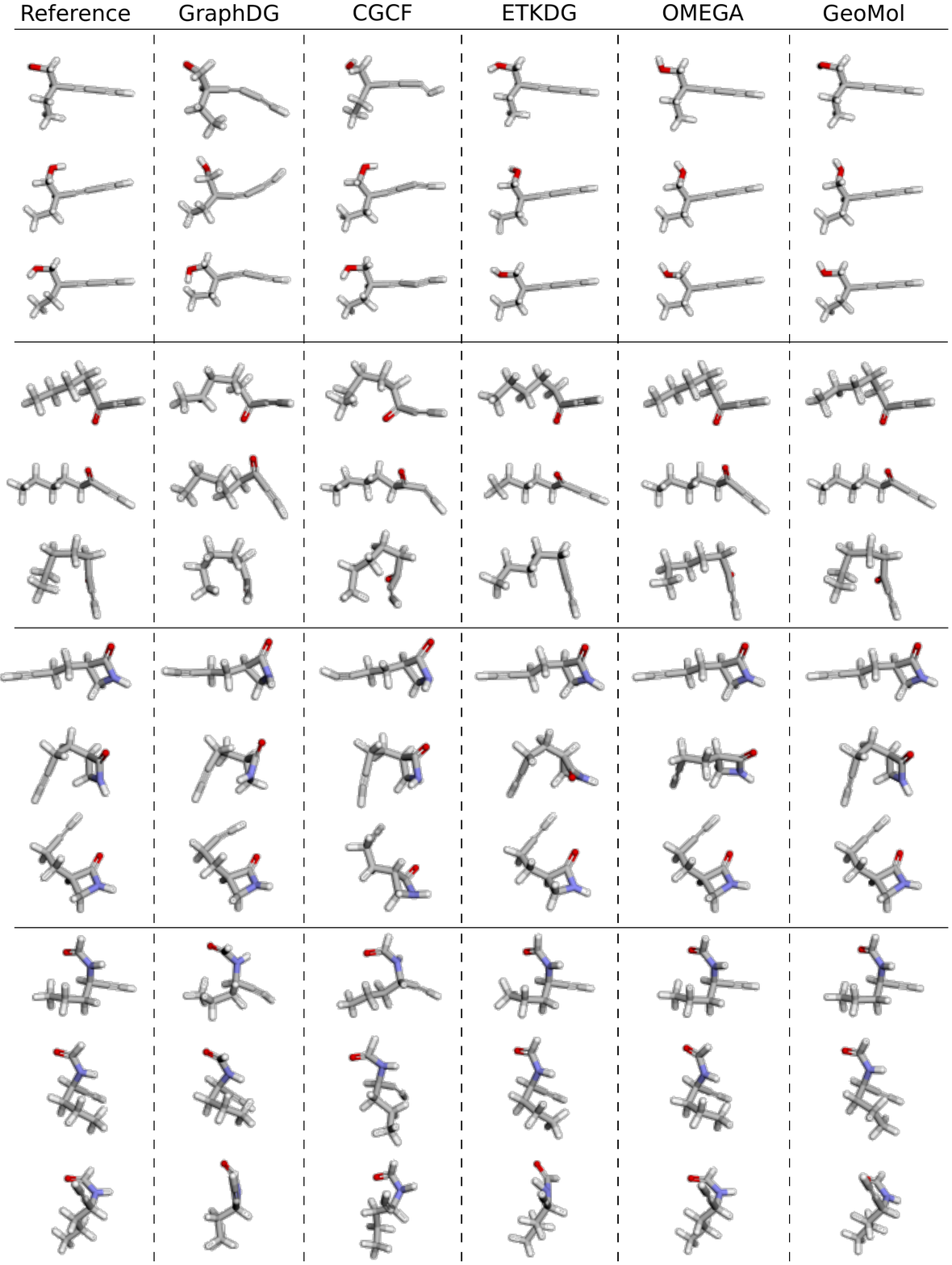}
    \caption{Additional examples of generated conformers for the GEOM-QM9 dataset.}
    \label{fig:more_qm9}
\end{figure}

\begin{figure}
    \includegraphics[width=1\linewidth]{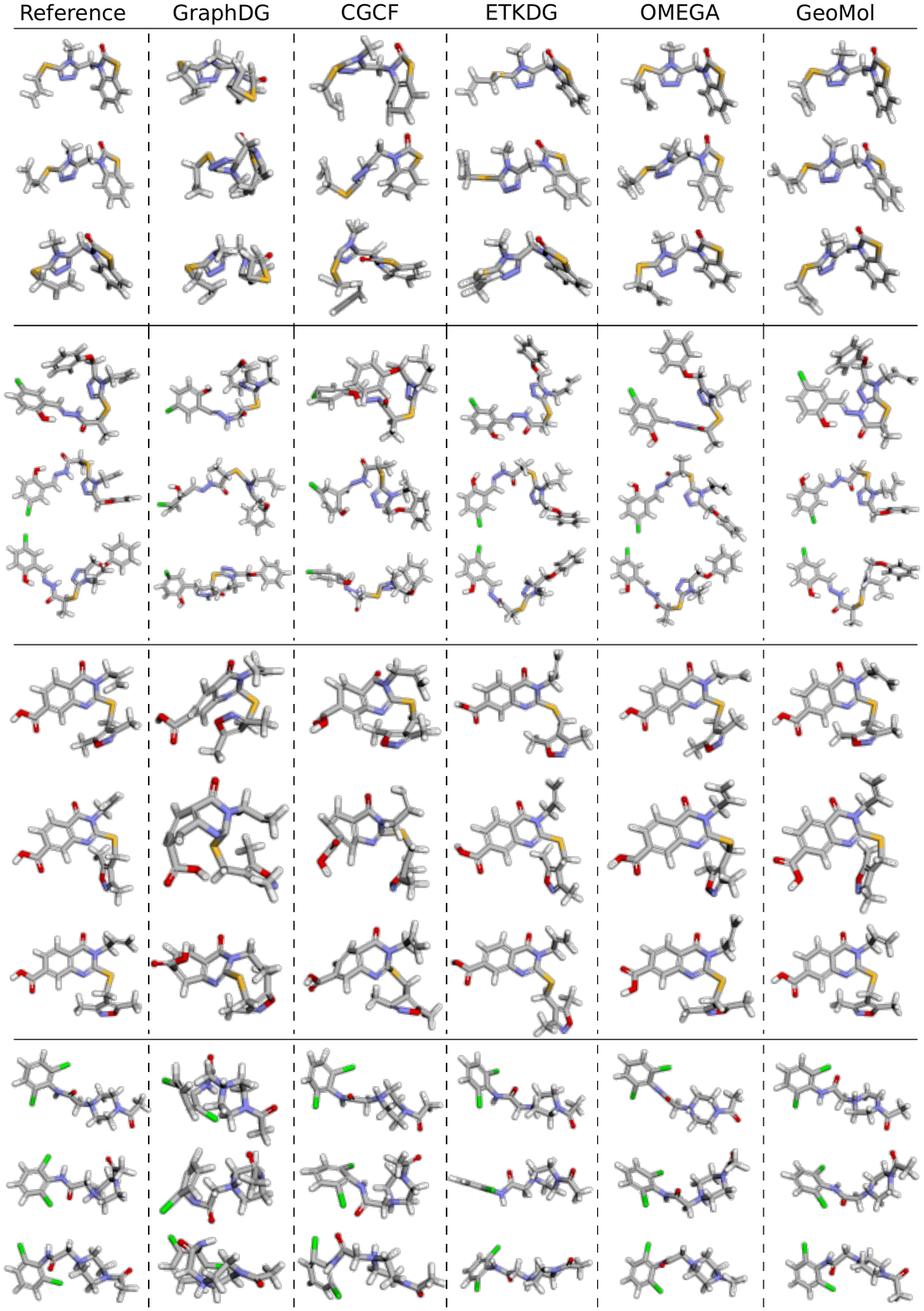}
    \caption{Additional examples of generated conformers for the GEOM-DRUGS dataset.}
    \label{fig:more_drugs}
\end{figure}

We show additional examples of generated conformers from the GEOM-QM9 test set in \cref{fig:more_qm9} and from the GEOM-DRUGS test set in \cref{fig:more_drugs}.

\section{Additional discussion}\label{apx:add_discussion}

\subsection{Purpose of conformer generation}
The purpose of conformer generation in the context of this work warrants some discussion. As stated in the main text, this purpose is heavily application-dependent. 
The current method is very helpful for computing the properties of drug-like molecules at low temperature, where it is necessary to generate a set of conformers that represents those with high Boltzmann populations. It can also be used for small molecules, but if the molecules are very small, it may be possible to identify all the important conformers by exhaustive search. At very high temperatures, where a large number of conformers are energetically accessible, it may be more convenient to represent the potential energy surface as a sum of hindered rotor potentials. 
Yet another common application for 3D conformer generation is the discovery of docked poses and 4D QSAR, where speed of the conformer generator becomes a crucial factor for virtually screening millions of structures. Crystal structure prediction (CSP) similarly relies on screening stable conformations, with recent models placing a greater emphasis on quality over quantity. These last two applications require modelling conformational flexibility rather than finding wells in the potential energy surface, as docked and crystallized conformations can fall outside these wells, carrying additional strain. The methods we present in this work can address any of these latter applications as well. We emphasize that the datasets on which \textsc{GeoMol} (or any ML model) is trained guide the downstream usage, but cleverly finetuning \textsc{GeoMol} on various downstream applications is an exciting future research direction.

\subsection{Comment on evaluation metrics}
Additionally, we note that while the evaluation metrics presented in this work are a sensible way to benchmark ML conformer generators against each other, they do not necessarily present a realistic evaluation of these algorithms in a production setting. This is why evaluating against the OMEGA software, which does not allow the user to request a specific number of conformers, is difficult. Further, for many applications, the generated conformers should be pruned and ranked (e.g., ranking conformers by energy and stable packing arrangements for CSP), and different conformer generation methods have varying methods to perform these post-processing steps. Because the scope of the presented work aims to distinguish our model against the many emerging ML-based conformer generators, we leave the exploration of such post-processing algorithms and strategies for future work. Furthermore, both the ETKDG and OMEGA algorithms are optimized to reproduce Protein Data Bank/Cambridge Structural Database conformations rather than replicate xTB low-energy conformations. In this sense, the evaluation presented here should not be viewed as a comprehensive argument for or against any of these methods.

\subsection{QM9 results}
Focusing on the QM9 results, we see both the power and drawbacks of the popular conformer generators such as OMEGA and ETKDG. Both methods can successfully generate diverse sets of conformers for the small molecules in the QM9 dataset, as shown by the high recall coverage values. However, both methods run into similar issues. That is, they both fail to generate the requested number of conformers for a few structures in the QM9 dataset, most of which have fused rings and multiple chiral centers specified. Since OMEGA, which constructs initial 3D structures from its database of fragments, does not have many of these fragments readily available, both OMEGA and ETKDG likely run into the same issue: using DG-based methods to generate initial 3D conformers for molecules with multiple chiral centers is difficult because of the implicit chiral constraints that must be satisfied. Take for example the following SMILES, which represents a fairly simple fused ring structure: 'C\#C[C@@]1(C)[C@H]2C[C@]1(C)C2'. ETKDG fails to embed a single conformer for this structure with repeated attempts. Removing the chiral centers specifications (i.e. inputting 'C\#CC1(C)C2CC1(C)C2' as the SMILES) results in a successful structure generation, but many of the outputs have the incorrect chirality. While these can sometimes be corrected with energy-based minimization, the key issue of determining chiral centers exactly with DG-based methods remains. \textsc{GeoMol} specifically tackles this problem such that all specified chiral centers are solved exactly (see \cref{ssec:ls}).

\subsection{DRUGS results}
The results for the DRUGS half of GEOM dataset suggest additional issues with traditional conformer generation methods. With these larger, drug-like molecules with many rotatable bonds, the search space of conformers grows exponentially. Here, it is instructive to separate the performance of ETKDG from OMEGA. ETKDG relies on randomly sampling the space of possible interatomic distances, which leads to poor coverage for conformers where this space is large. If the goal is to improve coverage, ETKDG could benefit from improved sampling strategies. OMEGA systematically varies all torsion angles in the structure to generate ensembles, but since this process can become prohibitively expensive, we must include an RMSD cutoff for many of these input molecules so as to return results in a reasonable period of time. Furthermore, not only does OMEGA uses a modified MMFF94 FF to generate its fragment library, it also uses a FF to score each potential conformer; its identified low-energy structures will be different than the xTB geometries we use as ground truths. This is a limitation with the OMEGA algorithm as currently constructed, and it could benefit from higher quality fragment structures and improved methods to score output conformers. This is not to understate the power of OMEGA as a conformer generation tool; it significantly outperforms the other baselines in our study and should be viewed as state-of-the-art in reference.

\subsection{Runtime comparison}
While the baselines ML methods have not benchmarked the speed of their algorithms, we believe this is a crucial evaluation of any conformer generator, especially if the intent is to use such a model for pose prediction and high throughput virtual screening. For each baseline, we request a fixed number of conformers and record the time necessary to output a single conformer, parallelizing each model over 40 CPU cores. We repeat this analysis three times for each method and plot the results in \cref{fig:examples}. Importantly, we do not include the start up times for any of the methods (i.e. loading the neural network weights for the ML models). OMEGA, GraphDG, and \textsc{GeoMol} show comparative performance, with ETKDG lagging slightly behind. Additionally, \textsc{GeoMol} scales well for structures with an increasing number of rotatable bonds, especially compared to GraphDG and ETKDG. The CGCF model is orders of magnitude slower than the other baselines, which restricts its utility for high throughput applications of conformer generation.

\subsection{Limitations} \label{apx:limitations}
\begin{figure}
    \begin{center}
    \includegraphics[width=0.6\linewidth]{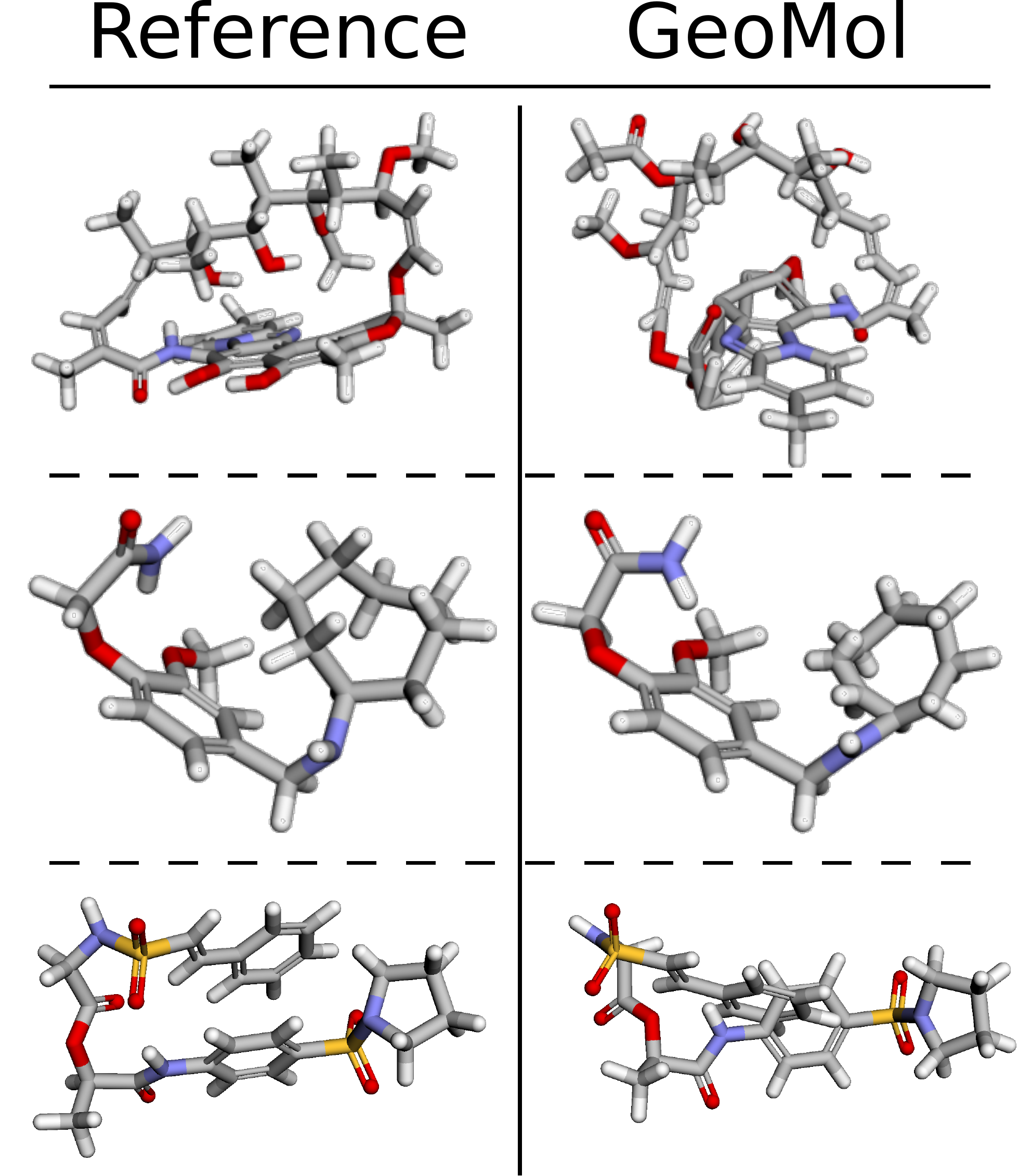}
    \end{center}
    \caption{Common \textsc{GeoMol} errors. Top: example of a macrocyle predicted poorly by \textsc{GeoMol}.  Middle: example of a larger ring with few training set occurrences. Bottom: example of a steric clash due to weak modeling of long-range interactions.}
    \label{fig:failures}
\end{figure}

The most significant drawback of our model is its weakness in capturing long-range interactions, which can manifest itself in several ways, detailed in \cref{fig:failures}. First, \textsc{GeoMol} poorly models macrocycles. Because of the formulation we use to construct conformers, error accumulation sometimes leads to imperfect structures, as the ring smoothing algorithm breaks down for large cycles (top of \cref{fig:failures}). We also qualitatively observe that large rings with very sparse examples in the training data can be predicted poorly by our model at test time. The middle of \cref{fig:failures} shows one such example of an eight-membered ring; here, additional training data should help, as there only exist a few eight-membered rings in the training set. Finally, the presence of steric clashes is an important issue with the current model. The bottom of \cref{fig:failures} shows one such example, being the most common error mode for \textsc{GeoMol}. Since global conformer information during torsion predictions is only conveyed through the graph network representations, it is challenging to account for the position of atoms far away in graph connectivity, which can lead to overlapping atoms in the final prediction. This is especially troubling for interactions such as the pi-pi stacking depicted in the figure, which require moieties in the conformer with large 2D distances in the molecular graph to be predicted close to one another in 3D space. For this reason, extending \textsc{GeoMol} with an improved method for long-range interactions is an exciting future direction.

\subsection{Conclusions}
In theory, \textsc{GeoMol} should replicate the quality of conformers on which it was trained, which gives the model unique power and flexibility for downstream applications. For example, one could easily construct a smaller database of high-quality conformers optimized at a higher level of theory than semi-empirical (ex. coupled-cluster) and finetune \textsc{GeoMol}. These higher-quality conformers could even include solvent effects in a range of solvents, which are often neglected in traditional conformer generation methods. Indeed, these downstream modifications could have significant impacts on a range of applications in cheminformatics and drug discovery.

Both the GraphDG and CGCF models, which represent important steps in ML-enabled conformer generation, are not quite ready to be used in a production setting without relying on a force field optimization. This is especially evident given the example structures presented in the main text and in \cref{apx:more_ex}. However, both models improve over traditional baselines at searching the space of interatomic distances and outputting diverse predictions, which is a credit to the search strategies employed by each model, especially CGCF. Further, the GraphDG model is competitive with \textsc{GeoMol} for the QM9 dataset and should be viewed as a viable alternative to \textsc{GeoMol} for small molecules with few heavy atoms.

As researchers continue to develop 3D ML-based conformer generators, the discussion between DG-based and non DG-based algorithms will carry on. Here, we chose to take the latter approach because we wanted to explicitly model important elements of conformer generation, and this strategy has been successful. In our experience, modeling these specific geometric elements leads to a significant improvement in the quality of the output structures. It is still unclear if such an approach can tackle long-range interactions, which remains the major drawback of our model. We hope to answer this question in future work.